\documentclass[twocolumn]{aastex631}
\pdfoutput=1

\newcommand{\MalskyEtAlP}{\citep{MalskyEtAl2025}}
\newcommand{\MalskyEtAlT}{\citet{MalskyEtAl2025} }

\received{Nov 4, 2024}
\revised{March 28, 2025}

\submitjournal{ApJ}

\begin{document}

\title{The radiative effects of photochemical hazes on the atmospheric circulation and phase curves of sub-Neptunes}

\author[0000-0001-8342-1895]{Maria E. Steinrueck}
\altaffiliation{51 Pegasi b fellow}
\affiliation{Department of Astronomy \& Astrophysics, University of Chicago, Chicago, IL, USA}
\affiliation{Max-Planck-Institut f\"ur Astronomie, 69117 Heidelberg, Germany}

\author[0000-0001-9521-6258]{Vivien Parmentier}
\affiliation{Universit\'e C\^ote d’Azur,    Observatoire de la C\^ote d’Azur, CNRS, Laboratoire Lagrange, France}
\affiliation{Atmospheric, Ocean, and Planetary Physics, Department of Physics, Oxford University, OX1 3PU, UK}

\author[0000-0003-0514-1147]{Laura Kreidberg}
\affiliation{Max-Planck-Institut f\"ur Astronomie, 69117 Heidelberg, Germany}

\author[0000-0002-8518-9601]{Peter Gao}
\affiliation{Carnegie Science Earth \& Planets Laboratory, Washington, DC, USA}

\author[0000-0002-1337-9051]{Eliza M.-R. Kempton}
\affiliation{Department of Astronomy, University of Maryland, College Park, MD, USA}

\author[0000-0002-0659-1783]{Michael Zhang}
\altaffiliation{51 Pegasi b fellow}
\affiliation{Department of Astronomy \& Astrophysics, University of Chicago, Chicago, IL, USA}

\author[0000-0002-7352-7941]{Kevin B. Stevenson}
\affiliation{JHU Applied Physics Laboratory, 11100 Johns Hopkins Rd, Laurel, MD 20723, USA}

\author[0000-0003-0217-3880]{Isaac Malsky}
\affil{Jet Propulsion Laboratory, California Institute of Technology, Pasadena, CA 91109, USA}

\author[0000-0001-8206-2165]{Michael T.\ Roman}
\affiliation{School of Physics and Astronomy, University of Leicester, Leicester, UK}

\author[0000-0003-3963-9672]{Emily Rauscher}
\affil{Department of Astronomy and Astrophysics, University of Michigan, Ann Arbor, MI, 48109, USA}

\author[0000-0002-2110-6694]{Matej Malik}
\affiliation{Department of Astronomy, University of Maryland, College Park, MD, USA}

\author[0000-0003-3444-5908]{Roxana Lupu}
\affiliation{Eureka Scientific, Inc., Oakland, CA, USA}

\author[0000-0003-3759-9080]{Tiffany Kataria}
\affiliation{NASA Jet Propulsion Laboratory, California Institute of
Technology, Pasadena, CA, USA}

\author[0000-0002-4487-5533]{Anjali A. A. Piette}
\affiliation{School of Physics \& Astronomy, University of Birmingham, Edgbaston, Birmingham, UK}

\author[0000-0003-4733-6532]{Jacob L.\ Bean}
\affil{Department of Astronomy \& Astrophysics, University of Chicago, Chicago, IL USA}

\author[0000-0001-8236-5553]{Matthew C.\ Nixon}
\affiliation{Department of Astronomy, University of Maryland, College Park, MD, USA}




\begin{abstract}

Measuring the atmospheric composition of hazy sub-Neptunes like GJ~1214b through transmission spectroscopy is difficult because of the degeneracy between mean molecular weight and haziness. It has been proposed that phase curve observations can break this degeneracy because of the relationship between mean molecular weight (MMW) and phase curve amplitude. However, photochemical hazes can strongly affect phase curve amplitudes as well. We present a large set of GCM simulations of the sub-Neptune GJ~1214b that include photochemical hazes with varying atmospheric composition, haze opacity and haze optical properties. In our simulations, photochemical hazes cause temperature changes of up to 200~K, producing thermal inversions and cooling deeper regions. This results in increased phase curve amplitudes and adds a considerable scatter to the phase curve amplitude--metallicity relationship. However, we find that if the haze production rate is high enough to significantly alter the phase curve, the secondary eclipse spectrum will exhibit either emission features or strongly muted absorption features. Thus, the combination of a white-light phase curve and a secondary eclipse spectrum can successfully distinguish between a hazy, lower MMW and a clear, high MMW scenario. 

\end{abstract}



\section{Introduction} \label{sec:intro}
Sub-Neptunes are among the most common planets found around other stars \citep{HowardEtAl2012}. Given that no such planet exists in our own Solar System, planet formation models historically did not account for them. The formation \citep{FortneyEtAl2013, LeeChiang2016SuperEarthFormation,LambrechtsEtAl2019SuperEarthFormation, ZengEtAl2019GrowthModel, BurnEtAl2024RadiusValley}, evolution \citep{OwenWu2013,GinzburgEtAl2018CorePoweredMassLoss, Mordasini2020} and interior structure \citep[e.g.,][]{RogersEtAl2010GJ1214b, RogersEtAl2011ExoNeptuneFormation, NettelmannEtAl2011GJ1214bInterior, ValenciaEtAl2013GJ1214b, LuquePalle2022, NixonEtAl2024GJ1214b} of these planets are currently a subject of debate. Measuring the atmospheric composition of sub-Neptunes is expected to provide constraints on the history of sub-Neptunes and to break degeneracies in interior structure models  \citep[e.g.,][]{BeanEtAl2021SubNeptuneReview, NixonEtAl2024GJ1214b, BennekeEtAl2024TOI-270d}.

Transmission spectroscopy of sub-Neptunes has proven to be challenging due to the degeneracy between aerosols (photochemical hazes and condensate clouds) and mean molecular weight (MMW), both of which can reduce the amplitude of spectral features \citep{BennekeSeager2012, LineParmentier2016}. This degeneracy is particularly limiting for observations taken in a narrow bandpass such as the WFC3 G141 grism of the Hubble Space Telescope \citep [e.g.,][]{KreidbergEtAl2014, KnutsonEtAl2014HD97658b, CrossfieldKreidberg2017, BrandeEtAl2024}. The wider wavelength coverage and higher precision of JWST are able to mitigate this degeneracy in some cases due to covering multiple absorption bands of the same species and resolving subtle differences in spectral shape between those two scenarios \citep[e.g.,][]{BennekeSeager2012, BennekeSeager2013, MaiLine2019, Piaulet-GhorayebEtAl2024}. Yet, aerosols remain a challenge even with JWST \citep[e.g.,][]{WallackEtAl2024TOI-836c}. Furthermore, the benchmark sub-Neptune GJ 1214b, which has the highest transmission spectra metric \citep{KemptonEtAl2018TSM} of all known sub-Neptunes and thus should be the easiest target for atmospheric characterization, has an unexpectedly flat transmission spectrum \citep{KreidbergEtAl2014, GaoEtAl2023GJ1214b, SchlawinEtAl2024GJ1214b, OhnoEtAl2024GJ1214b}. The extremely low feature amplitude of its spectrum cannot be explained with a high mean-molecular weight alone and therefore requires high-altitude aerosols. 

Most sub-Neptunes that are promising targets for atmospheric characterization are in the temperature range of 300~K$<T_{eq}<$900~K. For these temperatures, models predict that photochemical hazes are likely to form \citep{MorleyEtAl2013, MorleyEtAl2015, KawashimaIkoma2018, KawashimaIkoma2019, LavvasEtAl2019, AdamsEtAl2019} and to be the most important opacity source in the atmosphere \citep{GaoEtAl2020}. Laboratory experiments also suggest that photochemical hazes readily form across a range of atmospheric compositions and temperatures relevant to sub-Neptunes \citep{HorstEtAl2018,HeEtAl2018, HeEtAl2020}. For GJ 1214b in particular, general circulation models (GCMs)\citep{CharnayEtAl2015a} and microphysics models \citep{OhnoEtAl2018,OhnoEtAl2020FluffyAggregates} suggest that it may be hard to lift condensate clouds to high enough altitudes to explain the flatness of the spectrum. Thus, photochemical hazes have become a favored explanation for the featureless atmosphere of GJ 1214b in particular \citep[e.g.,][]{GaoEtAl2023GJ1214b}.

Phase curves and emission spectra are alternative ways of characterizing the atmospheres of sub-Neptunes that could be used to break the degeneracies from transmission spectroscopy. In particular, phase curves have been proposed as an alternative measurement of the mean molecular weight of sub-Neptune atmospheres \citep{JWSTCycle1GOBeanGJ1214bPhaseCurve}. This is because atmospheric dynamics predicts a relationship between phase curve amplitude and mean molecular weight, with an increase in the phase curve amplitude with increasing mean molecular weight \citep{KatariaEtAl2014,ZhangShowman2017}.  \citet{KemptonEtAl2023} recently applied this approach to the benchmark sub-Neptune GJ~1214b, utilizing JWST MIRI LRS to observe the first thermal phase curve of a sub-Neptune. Emission spectra are also of interest for planets with strong aerosol coverage because they probe deeper into the atmosphere than transmission spectra \citep{FortneyEtAl2005Transmission}.

However, 1D models show that photochemical hazes can strongly impact the temperature structure  \citep{MorleyEtAl2015, PietteMadhusudhan2020MiniNeptunes, LavvasArfaux2021}. The absorption of stellar photons by hazes results in a thermal inversion at low pressures and a cooling of the deep atmosphere.  3D simulations of hot Jupiters show that for sufficiently high haze production rates, hazes can strongly affect atmospheric circulation, increasing or decreasing the strength of the equatorial jet depending on the assumed haze refractive index, and increase the phase curve amplitude \citep{SteinrueckEtAl2023}. It thus seems likely that the mean molecular weight-phase curve amplitude relation for sub-Neptunes, which so far has been mainly studied under the assumption of clear atmospheres \citep[with one study including radiative feedback of condensate clouds, ][]{CharnayEtAl2015b} will be affected by photochemical hazes. Yet, the radiative feedback of hazes in GCMs of sub-Neptunes has not been studied systematically.

While condensate clouds in sub-Neptune atmospheres have been studied using GCMs with multiple approaches \citep{CharnayEtAl2015a,CharnayEtAl2015b,ChristieEtAl2022}, photochemical hazes have not been included in sub-Neptune GCMs until recently. 
As part of the effort to interpret the phase curve observations of GJ~1214b, we included a large grid of GCM simulations of that planet with horizontally homogeneous hazes in the observational paper \citep{KemptonEtAl2023}. These were the first published sub-Neptune GCMs that considered the radiative effects of a  haze layer. The models showed that hazes can have a first-order effect on the phase curve and that both a high metallicity and highly scattering aerosols were required to match the observed phase curve.
In this publication, we are conducting an in-depth analysis of this large and rich set of simulations of GJ~1214b to better understand how hazes affect the atmospheres of sub-Neptunes and what conclusions can be drawn for the broader sub-Neptune population.

The remainder of this paper is structured as follows: In Section \ref{sec:methods}, we describe our modeling approach. Section \ref{sec:results} describes temperature structure and atmospheric circulation of the simulations, followed by Section \ref{sec:observabletrends} focused on observable trends in the 5-12~$\mu$m region. Further, we briefly discuss how our models compare to the recent JWST MIRI LRS observations of GJ~1214b, including the original \citep{KemptonEtAl2023} data reduction and an alternative data reduction presented in a second paper from our collaboration \MalskyEtAlP, in Section \ref{sec:gj1214b_comparison}.  Finally, we discuss possible observing strategies for future observations of sub-Neptunes in Section \ref{subsec:observingstrategies} and summarize our conclusions in Section \ref{sec:conclusion}.

\section{Methods}
\label{sec:methods}
\begin{deluxetable}{lcc}
\tablecaption{Model parameters shared by all simulations \label{tab:modelparameters}}
\tablehead{\colhead{Parameter} & \colhead{Value} & \colhead{Units}}
\startdata
Radius\tablenotemark{1} & $1.7469282 \times 10^7$ & m \\
Gravity\tablenotemark{1} & 10.65 & m s$^{-2}$ \\
Rotation period\tablenotemark{1, 2}& 1.58 & d \\
Semimajor axis\tablenotemark{1} & 0.0149 & AU \\
Interior flux\tablenotemark{3} & $4.59 \times 10^{-2}$ & W m$^{-2}$ \\
Horizontal resolution & C32\tablenotemark{a} & \\
Vertical resolution\tablenotemark{4} & 60 & layers \\
Lower pressure boundary\tablenotemark{4} & $1.75 \times 10^{-7}$ & bar \\
Upper pressure boundary & 200 & bar \\
\enddata
\tablenotetext{1}{\citet{CloutierEtAl2021}}
\tablenotetext{2}{assumed to be equal to the orbital period due to tidal locking}
\tablenotetext{3}{Corresponding to an intrinsic temperature of 30~K, based on \citet{LopezFortney2014}}
\tablenotetext{4}{With exception of the 3,000$\times$ solar clear-atmosphere simulation, which has 53 layers and a lower pressure boundary of $2 \times 10^{-6}$~bar}
\tablenotetext{a}{Equivalent to a resolution of 128x64 on a longitude-latitude grid}
\end{deluxetable}

\begin{deluxetable*}{lrrrrr}
\tablecaption{Composition-dependent model parameters \label{tab:composition-dependent_modelparameters}}
\tablehead{
\colhead{Composition} & \colhead{Mean molecular weight} & \colhead{Specific heat capacity} & \colhead{Specific gas constant} & \colhead{Hydrodynamic time step} & \colhead{Radiative time step} \\
\colhead{} & \colhead{[g mol$^{-1}]$} & \colhead{[J kg$^{-1}$ K$^{-1}$]} & \colhead{[J kg$^{-1}$ K$^{-1}$]} & \colhead{[s]} & \colhead{[s]}
}
\startdata
1$\times$ solar & 2.24 & 13,000 & 3714 & 25 & 50 \\
100$\times$ solar & 4.38 & 6,474 & 1,898 & 25 & 50 \\
300$\times$ solar & 8.50 & 3,500 & 978 & 10 & 20 \\
3,000$\times$ solar & 28.5 & 1,300 & 292 & 10 & 20 \\
\enddata
\end{deluxetable*}

We use SPARC/MITgcm \citep{ShowmanEtAl2009,KatariaEtAl2013}, which couples the dynamical core of \citet{AdcroftEtAl2004} to the wavelength-dependent radiative transfer code of \citet{MarleyMcKay1999}, to simulate the atmosphere of sub-Neptune GJ~1214b. SPARC/MITgcm has in the past mainly been used to simulate the atmospheres of hot Jupiters \citep[e.g.,][]{ParmentierEtAl2013,KatariaEtAl2016,SteinrueckEtAl2019,SteinrueckEtAl2023,TsaiSteinrueckEtAl2022}, but has also been successfully applied to hot Neptunes \citep{LewisEtAl2010} and sub-Neptunes \citep{KatariaEtAl2014}.

\subsection{Atmospheric Dynamics}
\label{sec:methods_dynamics} 
We use the dynamical core of \citet{AdcroftEtAl2004} to solve the primitive equations on a cubed-sphere grid. Our numerical setup uses a fourth-order Shapiro filter \citep{Shapiro1970} to damp numerical fluctuations at the grid scale that could otherwise lead to instabilities. Further, we include a pressure-dependent linear drag, $( d\mathbf{v}/dt)_{\textrm{drag}} = - k_v \mathbf{v}$, in the deep atmosphere ($p > p_{\textrm{drag,top}} = 10$~bar), which ensures convergence independent of initial conditions given long enough simulation run times \citep{LiuShowman2013} and stabilizes the simulation. The drag strength is given by $k_v=k_F (p - p_{\textrm{drag,top}})/(p_{\textrm{bottom}}-p_{\textrm{drag,top}})$, with the bottom boundary of the simulation $p_{\textrm{bottom}}=200$~bar and the maximum drag strength $k_F = 10^{-5}$~s$^{-1}$. The most important model parameters are summarized in Table \ref{tab:modelparameters} for parameters shared by all simulations and Table \ref{tab:composition-dependent_modelparameters} for parameters depending on the assumed composition of the atmosphere.

\subsection{Radiative Transfer}
\label{sec:methods_radtran}
The radiative transfer code used in our simulations \citep{MarleyMcKay1999} assumes a plane-parallel atmosphere and uses the two-stream approximation.
Wavelength dependence is taken into account with the correlated-k method with 11 wavelength bins \citep{KatariaEtAl2013} within the simulations and 196 wavelength bins for post-processing simulations to obtain spectra.
We used molecular opacities based on \citet{FreedmanEtAl2008, FreedmanEtAl2014} and \citet{LupuEtAl2014}, with the more recent updates described in \citet{MarleyEtAl2021Sonora}. 
For the opacity calculation, atmospheric abundances at thermochemical equilibrium were calculated over a grid of temperatures (75-6000 K) and pressures (10$^{-6}$ to 3000 bar) based on the models in \citet{Gharib-NezhadEtAl2021} and the Sonora atmosphere model \citep{MarleyEtAl2021Sonora}, extended here to metallicities up to 3000$\times$~solar. While the atmosphere of GJ~1214b is not expected to be in chemical equilibrium, studies show that for planets in the temperature range of GJ~1214b, at least for hydrogen-dominated atmospheres, the impact of transport-induced disequilibrium chemistry on the temperature structure is smaller than the possible effects of hazes \citep{ZamyatinaEtAl2023,SteinrueckEtAl2019}. Furthermore, the current data have so far been insensitive to disequilibrium species.

For the incoming starlight, we used a stellar spectrum based on a PHOENIX model \citep{HusserEtAl2013PHOENIX} using the stellar parameters of GJ~1214b from \citet{CloutierEtAl2021}.
At the bottom boundary of the model, we include an upward radiative flux corresponding to an intrinsic temperature of 30~K \citep{LopezFortney2014}.

The white-light phase curves were calculated by integrating the planetary flux weighted by the MIRI filter response at each phase and dividing it by the similarly integrated stellar flux, following Eq. (1) of \citet{CharbonneauEtAl2005ThermalEmission}. Details about our post-processing procedure can be found in \citep{ParmentierEtAl2016}.

\subsection{Haze model}
\begin{figure*}
\begin{center}
\plotone{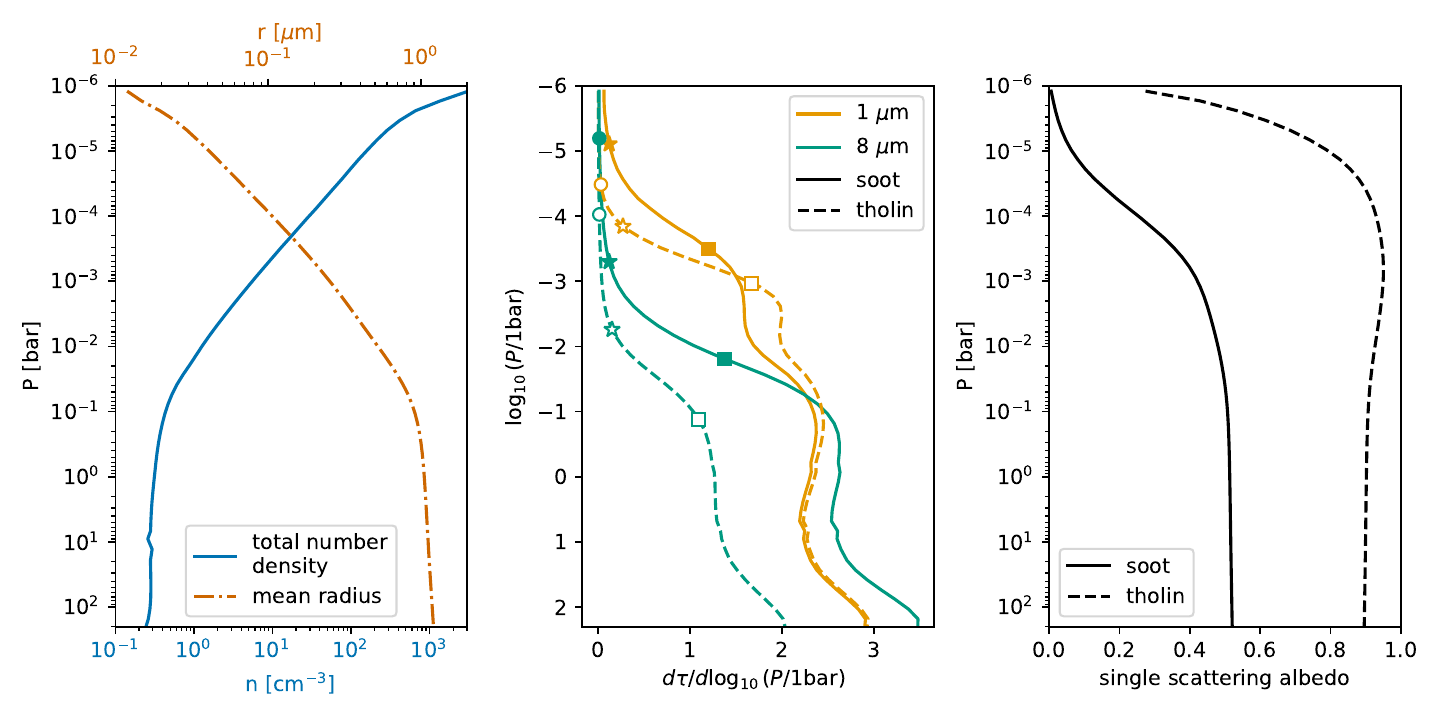}
\caption{Summary of the haze properties assumed in the simulations. The left panel shows number density and mean radius of the haze distribution that was used as input in the GCM simulations. The middle panel shows the differential optical depth for soot (solid) and tholin (dashed) hazes at 1 $\mu$m (representative for the incoming starlight) and 8 $\mu$m (representative for the outgoing thermal radiation) for the nominal haze opacity case. The pressure levels where the optical depth reaches 0.01 (circle), 0.1 (star), and 1 (square) are highlighted. For soot at 1~$\mu$m, the optical depth reaches 0.01 at a pressure of $6\cdot10^{-7}$~bar, outside the pressure range shown.  The right panel shows the wavelength-averaged single scattering albedo weighted by the stellar flux. For the ``maximally reflective'' case, the optical depth is identical to the soot case, but the single scattering albedo is fixed to 0.9999.}
\label{fig:hazeproperties_vs_pressure}
\end{center}
\end{figure*}
We assume a haze profile that is static and horizontally uniform on isobars. For a first exploration of the radiative effects of photochemical hazes in sub-Neptunes, this is justified because a tracer-based  model with radiatively active hazes as in \citet{SteinrueckEtAl2023} is much more computationally expensive and would not allow an as thorough exploration of the parameter space. Furthermore, for GJ~1214b-like conditions, haze particle radii are expected to vertically vary by more than two orders of magnitude \citep{KawashimaIkoma2019,LavvasEtAl2019}. It is thus likely that the vertical variations in particle radius, and thus opacity, are  larger than the horizontal ones.

The nominal haze profile was derived from a 1D simulation with CARMA \citep{TurcoEtAl1979CARMA,ToonEtAl1979CARMA,AckermanEtAl1995CARMA}, a bin-based microphysical model for clouds and hazes. In particular, the simulation setup of \citet{AdamsEtAl2019} with a haze production rate of $10^{-12}$~g~cm$^{-2}$s$^{-1}$, a metallicity of 100$\times$ solar and spherical hazes was used. This haze production rate is in the typical range of predicted haze formation rates for GJ~1214b based on photolysis mass fluxes of haze precursor molecules in photochemical models \citep{LavvasEtAl2019, KawashimaIkoma2019}. We choose the 100$\times$~solar case as it is in the intermediate range of metallicities that we consider. Throughout this work, we use the same haze profile independent of metallicity. We do not scale haze opacity as a function of metallicity, because neither experimental nor theoretical studies show evidence for a monotonous relationship between haze production rate and metallicity \citep{HorstEtAl2018, HeEtAl2018, LavvasEtAl2019}.   Number density and mean radius from the microphysics model are shown in the left panel of Fig. \ref{fig:hazeproperties_vs_pressure}.

We then calculated haze optical depth, single scattering albedo and asymmetry parameter as a function of pressure for the particle size distribution from the microphysics model with a Mie scattering code \citep{BohrenHuffmanAbsorptionScatteringBook}, assuming the same pressure grid as used in the microphysics model. Because the haze opacity varies weakly within each bin, we consider it as a continuum opacity that can just be added to the k-coefficients within each bin. We took the wavelength-average of each of these three quantities within each of the 11 wavelength bins used in the GCM and interpolated the profiles onto the vertical grid of the GCM. For the few GCM layers above the low-pressure boundary of the microphysics model (1 $\mu$bar), quantities were extrapolated linearly in pressure. 

To simulate different haze production rates, we scaled the optical depth profile by a factor of 10 or 100, while leaving the single scattering albedo and asymmetry parameter unchanged. This approach neglects that particle sizes can vary between different metallicities and haze production rates.
However, it provides a simple and fast way of exploring parameter space. It also makes it easier to isolate changes in atmospheric dynamics and radiative transfer due to the gas-phase composition. Multiple studies \citep{LavvasEtAl2019, KawashimaIkoma2019, AdamsEtAl2019, GaoEtAl2023GJ1214b} have explored the effects of varying metallicity and haze production rate on particle sizes in 1D. Our study aims to complement this work by focusing on changes in atmospheric dynamics while leaving the particle size distribution constant.

For the optical properties of the photochemical hazes, we explored three different scenarios. In the first scenario, we assumed refractive indices of soot based on the compilation of \citet{LavvasKoskinen2017} in the Mie calculation. This is a frequently used assumption in simulations of hot exoplanet atmospheres \citep[e.g.][]{MorleyEtAl2015,LavvasEtAl2019,AdamsEtAl2019,SteinrueckEtAl2021}, as soots can survive much higher temperatures than most other candidate haze compositions. The second scenario assumes refractive indices of Titan-like tholins \citep{KhareEtAl1984} as an example of more scattering hazes with a stronger wavelength dependence. As a third scenario, motivated by the observed high albedo of GJ~1214b, we also consider hazes with the extinction cross section and asymmetry parameter equal to the soot scenario but with the single scattering albedo set to 0.9999. We dub this scenario ``maximally reflective hazes''. These assumptions are identical to what was for ``maximally reflective hazes'' assumed in \citet{KemptonEtAl2023}. We note that compared to scattering and absorption calculated self-consistently from a realistic refractive index, this approach may overestimate extinction at long wavelengths, due to the absorption cross section dropping off $\propto \lambda^{-1}$ while the scattering cross section is $\propto \lambda^{-4}$ in the Rayleigh-scattering regime. However, our goal here is to provide an extreme limiting case for illustration, and this caveat is unlikely to qualitatively change the conclusions.

The optical depth profiles for two representative wavelengths and the single-scattering albedo profile averaged across the full wavelength range, weighed by the stellar spectrum are shown in the middle and right-hand panels of Fig. \ref{fig:hazeproperties_vs_pressure}. Furthermore, Fig. \ref{fig:hazeprops_vs_wavelength} illustrates the wavelength dependence of these properties at a range of pressures.

Recently, additional optical properties from experiments specifically aiming to simulate haze formation on exoplanets were published \citep{HeEtAl2024WaterRichOpticalProperties,CorralesGavilanEtAl2023}. These sets of optical constants still are qualitatively similar to the \citet{KhareEtAl1984} tholins in that they feature a strong UV opacity slope, an opacity window region with a very low extinction coefficient in the optical to near-infrared, moderate extinction coefficients at near-infrared wavelengths longer than the opacity window, and somewhat higher extinction coefficients in the mid-infrared. It thus seems likely that these optical property sets will produce qualitatively similar results to the tholin case, and we thus do not run separate simulations for them.

\subsection{Simulations}
Simulations were initiated from a state of rest, with a globally uniform initial pressure-temperature profile derived from a 1D radiative transfer calculation with HELIOS \citep{MalikEtAl2017HELIOS, MalikEtAl2019HELIOS}. All simulations were run for 1,000 Earth days simulation time. The results presented in this paper are based on the time-average of the last 100 simulation days. We ran a total of 27 simulations. This includes all 24 simulations listed in Extend Data Table 3 in \citet{KemptonEtAl2023} as well as 3 additional simulations with tholins (300$\times$~solar, haze scale factor 100$\times$; 3,000$\times$~solar, haze scale factor 10$\times$; 3,000$\times$~solar, haze scale factor 100$\times$;).

\section{Results}
\label{sec:results}
\subsection{Temperature structure}
\label{subsec:temperaturestructure}
\begin{figure}
\begin{center}
\plotone{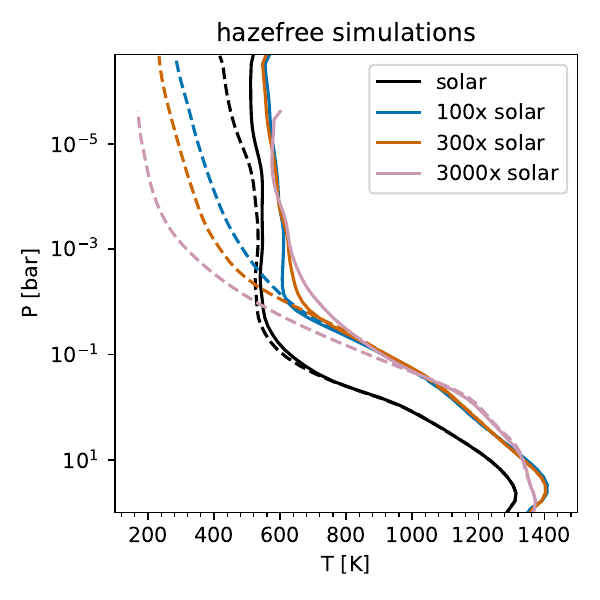}
\caption{Dayside-averaged (solid) and nightside-averaged (dashed) temperature profiles  for all clear-atmosphere simulations. Temperatures were weighted by the cosine of the angle of incidence (dayside)/the angle from the antistellar direction (nightside) during averaging. }
\label{fig:temperatureprofiles_hazefree}
\end{center}
\end{figure}

\begin{figure*}
\begin{center}
\plotone{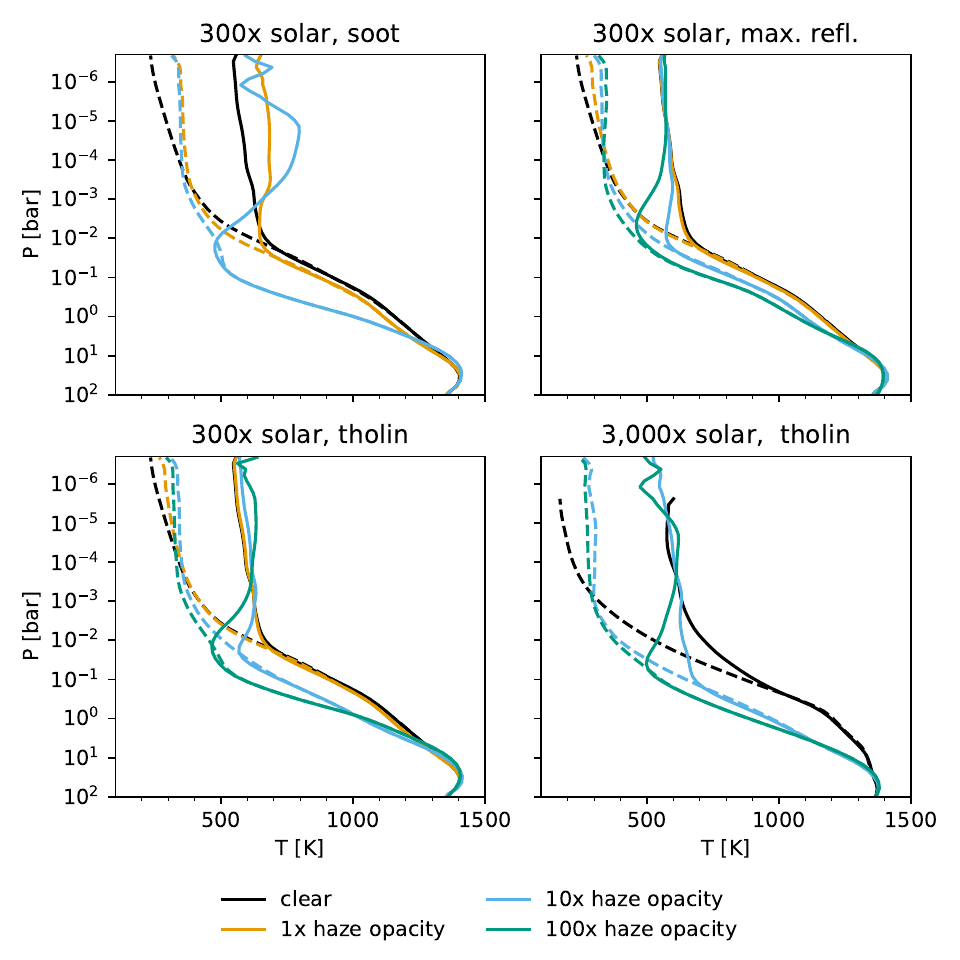}
\caption{Same as figure \ref{fig:temperatureprofiles_hazefree}  but for a selection of simulations with photochemical hazes compared to the clear simulation with the same metallicity. }
\label{fig:temperatureprofiles_hazy}
\end{center}
\end{figure*}
For clear-atmosphere models there is a well-established trend of less efficient heat redistribution and thus higher day-to-night temperature contrasts with higher mean molecular weight \citep{KatariaEtAl2014, CharnayEtAl2015a, ZhangShowman2017}. This behaviour is once again borne out by our clear-atmosphere models. (Fig. \ref{fig:temperatureprofiles_hazefree}). Introducing hazes can dramatically change the temperature structure of the atmosphere. The magnitude of these changes increases with enhanced haze opacity (Fig. \ref{fig:temperatureprofiles_hazy}). If the haze opacity increases sufficiently, all models form a dayside thermal inversion and exhibit a colder temperature profile in the deep atmosphere. However, the shape of the thermal inversion differs substantially depending on the assumed haze optical properties.

Soot hazes, being highly absorbing across the whole wavelength spectrum, lead to strong heating at low pressures, increasing dayside temperatures by up to 200~K at pressures $\lessapprox$1~mbar
At the same time, the atmosphere cools compared to the clear-atmosphere case at pressures $\gtrapprox$1--5~mbar due to the anti-greenhouse effect \citep[e.g., ][p. 449]{ThomasStamnes2002Book}. The strength of the thermal inversion strongly depends on the haze opacity: For the nominal haze opacity, the thermal inversion remains moderate, with a temperature increase of $\approx$40~K between 2 mbar and 0.3 mbar for the 300$\times$ solar case, with a large nearly isothermal region of the atmosphere above the inversion. For the 10$\times$ increased haze opacity, however, the thermal inversion spans over three orders of magnitude in pressure (between 20~mbar and 10 $\mu$bar for the $300\times$ solar case), with a temperature increase of over 300~K. This agrees well with the findings of \citet{MorleyEtAl2015}.

In the deep atmosphere ($p\gtrapprox0.1$~bar), where atmospheric heat transport is efficient, the nightside temperature profiles closely follow the dayside ones. Thus, nightside temperatures at high and intermediate pressures are cooler for higher haze opacities. This cooling effect is strongest for the soot hazes compared to other haze types. At very low pressures ($p\lessapprox0.1$~mbar), the nightside temperature profiles in the soot simulations become isothermal and are warmer than in the clear-atmosphere profile, which continues to decrease with height.

For the maximally reflective hazes, there is practically no absorption. Thus, the main way in which these hazes affect the temperature structure is by scattering incoming starlight at low pressures, preventing it from reaching deeper layers. The resulting dayside temperature profile is close to the haze-free temperature profile for the nominal haze opacity and exhibits a moderate-to-large thermal inversion for larger haze opacities. Due to the large amount of light reflected at low pressures, temperature profiles with maximally reflective hazes have cooler daysides at almost all pressures. However, compared to the soot case, the deeper region below the isothermal and inverted-temperature regions cools less.

Nightside temperatures in the deep atmosphere are again mainly set by the amount of dayside cooling. At low pressures ($p\lessapprox0.1$~mbar), the temperature profile becomes warmer and more isothermal with increasing haze opacity. This is likely due to a combination of radiative effects---due to the added haze opacity, radiation can escape less easily---and advection of heat from the dayside due to the increased strength of the equatorial jet at high haze opacities (see Section \ref{subsec:circulation}).

With a single-scattering albedo much higher than that of soot, but less than the maximally reflective case, the temperature changes in the tholin simulations exhibit some characteristics from each of these two extreme cases. The amount of cooling of deeper layers is comparable to or slightly higher than the maximally reflective case. Given the high single scattering albedo (reaching up to 0.95 between 0.1 and 1~mbar, see Fig. \ref{fig:hazeproperties_vs_pressure}), this appears to be mainly driven by reflection of incoming starlight. However, there are also small to moderate temperature increases at low pressures in all tholin simulations, owing to absorption by tholin haze. In most cases, the additional heating is not enough to change the structure of the atmosphere at low pressures (i.e., in the region above the cooling-driven thermal inversion, temperature still decreases with height). Only in simulations with 100$\times$ increased haze opacity does a more extended thermal inversion due to heating and a temperature peak at low pressures form that is comparable to the temperature structure in the soot simulations.
We further note that the temperature minimum due to cooling is located at higher pressures than in the maximally reflective case. This is likely caused by the reduced extinction cross section of tholins compared to soot and our maximally reflective hazes (for which the extinction cross section of soot was assumed) at low pressures (see Fig. \ref{fig:hazeproperties_vs_pressure} and \ref{fig:hazeprops_vs_wavelength}).

\subsection{Atmospheric circulation}
\label{subsec:circulation}
\begin{figure*}
\begin{center}
\plotone{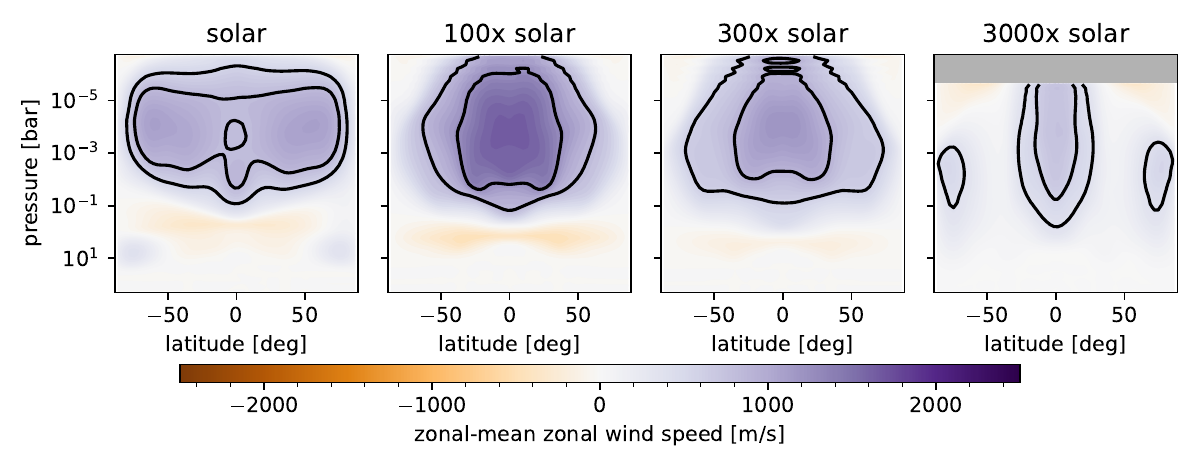}
\caption{Zonal-mean zonal velocity for the clear-atmosphere simulations. The contours highlight where the zonal-mean zonal velocity reaches 50\% and 75\% of the peak velocity within the simulation.}
\label{fig:uzonalav_clear}
\end{center}
\end{figure*}

The atmospheric circulation in the clear-atmosphere models is consistent with previous work and depends strongly on metallicity (Fig. \ref{fig:uzonalav_clear}). In the solar metallicity case, an equatorial jet as well as two mid-latitude jets (one per hemisphere) form, similar to previous simulations of GJ~1214b \citep{KatariaEtAl2014,CharnayEtAl2015a} and GJ~436b \citep{LewisEtAl2010}. The mid-latitude jets are broader and peak at lower pressures than the equatorial jet. The jets are not well-separated from each other---while there are distinct peaks, there are no regions of low zonal-mean zonal velocity or westward flow in between those peaks. In the 100$\times$~solar simulation, a single broad equatorial jet forms, resembling the results from \citep{CharnayEtAl2015a}. The zonal velocity pattern in the 300$\times$~solar case looks similar to the 100$\times$~solar case but the wind speeds are lower. This is consistent with the finding of \citet{ZhangShowman2017} that a higher mean molecular weight leads to lower wind speeds. The highest metallicity simulation, 3,000$\times$~solar, features three well-separated jets (one equatorial jet and one high-latitude jet in each hemisphere). \citet{ZhangShowman2017} suggest that a high mean-molecular weight leads to narrower jets (and thus a total higher number of jets on the planet for sufficiently high mean molecular weight) due to the smaller Rossby deformation radius. In their CO$_2$ and N$_2$ simulations using Newtonian cooling, high-latitude jets can be seen at similar locations as in our 3,000$\times$~solar simulation. However, the relative strength of their high-latitude jets compared to their equatorial jet is low (about 25\%), while in our simulations they reach more than 50\% of the equatorial jet speed, suggesting that full wavelength-dependent radiative transfer and detailed atmospheric composition matter. Qualitatively, the zonal-mean circulation is also in between the H$_2$O-rich and the 50\% H$_2$O-50\% CO$_2$ cases from \citet{KatariaEtAl2014} and also displays similarity to the H$_2$O simulation of \citet{CharnayEtAl2015a}.

\begin{figure*}
\begin{center}
\plotone{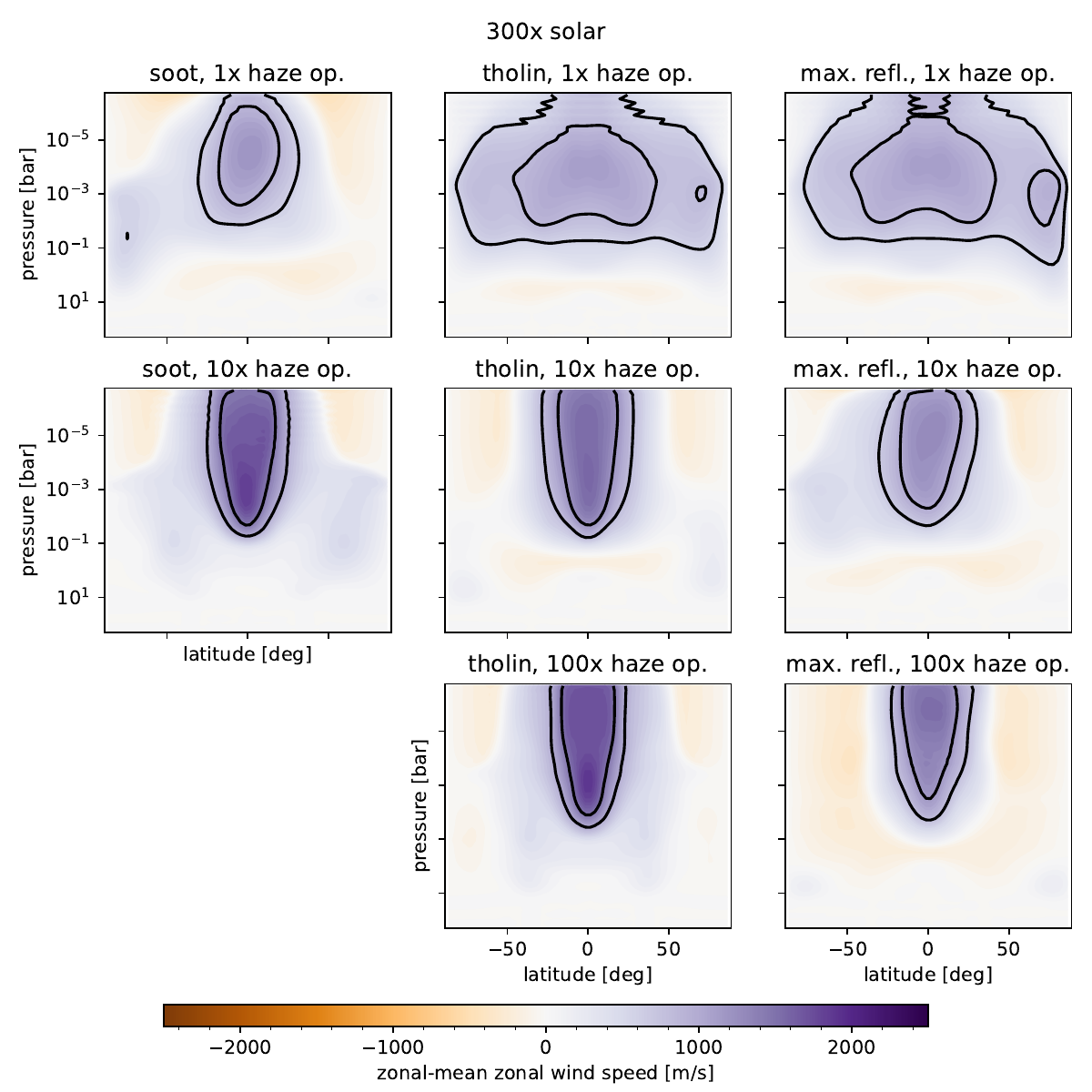}
\caption{Similar to Fig. \ref{fig:uzonalav_clear} but for all hazy simulations with a metallicity of 300$\times$~solar. The color scale is identical to Fig. \ref{fig:uzonalav_clear}.}
\label{fig:uzonalav_300xsolar}
\end{center}
\end{figure*}

\begin{figure*}
\begin{center}
\plotone{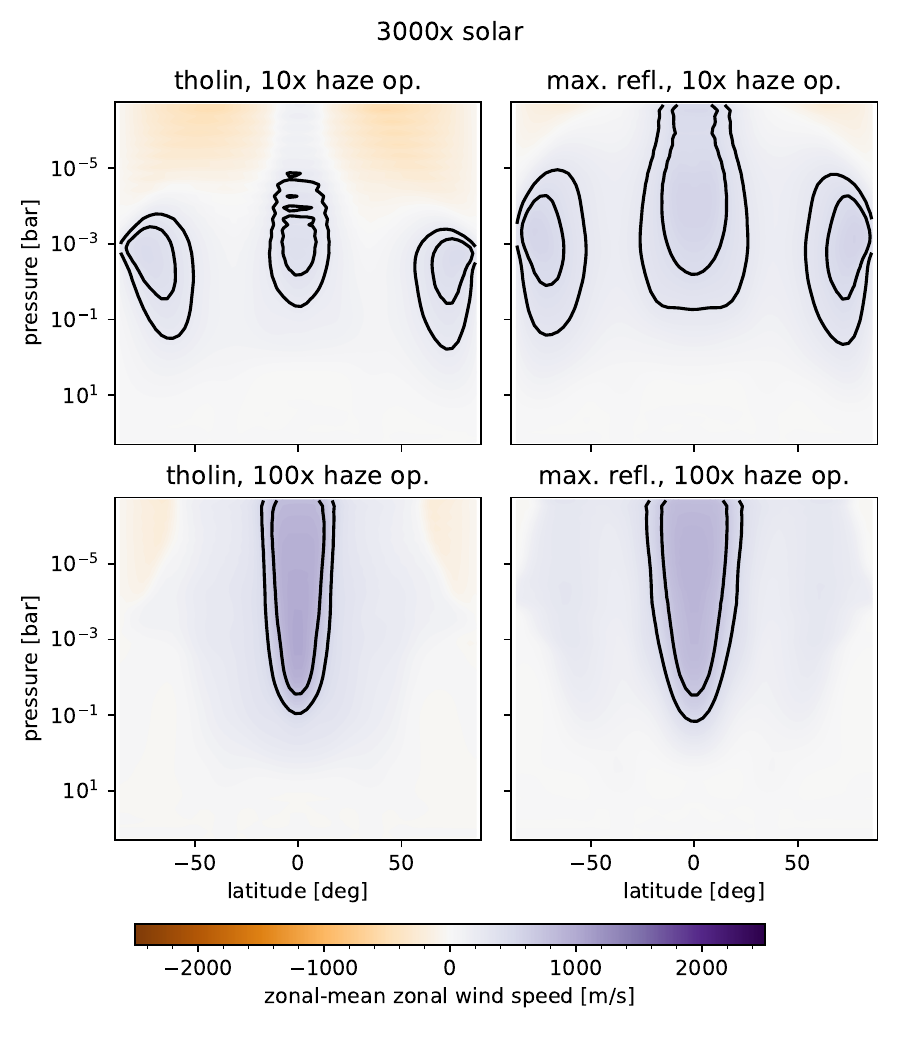}
\caption{Similar to Fig. \ref{fig:uzonalav_clear} but for hazy simulations with a metallicity of 3,000$\times$~solar. The color scale is identical to Fig. \ref{fig:uzonalav_clear}.}
\label{fig:uzonalav_3000xsolar}
\end{center}
\end{figure*}
For simulations including photochemical hazes, the atmospheric circulation transitions to a single super-rotating equatorial jet if the haze opacity is sufficiently high. Figure \ref{fig:uzonalav_300xsolar} illustrates this transition for the 300$\times$~solar case for all three different optical properties. A similar behavior can be observed for most other metallicities \footnote{The solar metallicity case shows a somewhat more complex behavior; however, the focus of our work is on higher metallicites.}. In most cases, the jet is relatively narrow, confined to $\pm$20 degrees of latitude. The jet speed typically increases with increasing haze opacity. For soot hazes, the transition to a narrow equatorial jet happens at a lower haze opacity than for tholins and maximally reflective hazes, likely because of their stronger impact on temperature structure and heating rates. 

The mechanism for the transition to a more focused and faster jet likely differs between the soot and the maximally reflective hazes. For the soot hazes, it seems likely that strongly enhanced thermal forcing in the regions of jet formation cause the transition. For the maximally reflective hazes, however, the net heating decreases due to the large fraction of radiation being reflected away from the planet. A different mechanism thus must cause the jet to  accelerate and narrow. One possibility is that due to the cooling of the atmosphere, the Rossby deformation radius shrinks and causes the jet to become confined to a smaller latitude range. The acceleration then might happen because there is a narrower latitude range at which the jet can transport heat from the dayside to the nightside. For the tholin case, both mechanisms could be at work. A detailed investigation of the mechanisms driving the atmospheric circulation changes, however, is outside the scope of our work.

The increased jet speed also leads to more efficient heat advection from the dayside to the nightside at low pressures. This is one of the factors leading to an increase in nightside temperatures compared to the haze-free case at very low pressures ($p\lessapprox10^{-4}$~bar, see Fig. \ref{fig:temperatureprofiles_hazy}). In the case of maximally reflective cases, the advection of heat leads to a somewhat decreased day-night temperature contrast at low pressures. In the soot case, however, the day-night temperature contrast at low pressures increases substantially, because the heating on the dayside greatly outweighs the increased temperature advection. For tholins, the day-night temperature difference decreases slightly in some simulations, while in other simulations it increases at some pressures and decreases at other pressures.

For higher metallicities, higher haze opacities are required for the atmospheric circulation to transition to a single jet. For example, for the 100$\times$~and 300$\times$~solar tholin and maximally reflective simulations, the single narrow jet already occurs at 10x the nominal haze opacity. For the 3000$\times$~solar case, the corresponding simulations features an equatorial jet as well as two high-latitude jets (Fig. \ref{fig:uzonalav_3000xsolar}). Only the simulations with 100x the nominal haze opacity have a single equatorial jet for this metallicity.
One possible explanation is that atmospheres with higher metallicity have a larger abundance of strongly absorbing gas species such as CO$_2$, H$_2$O and CO. Thus, the haze opacity makes up a lower fraction of the total opacity at the same haze opacity.

\section{Observable trends in the 5--12 micron region}
\label{sec:observabletrends}
Now we turn our attention to how the changes in temperature structure and atmospheric circulation described above translate to observable changes to phase curves and emission spectra. We focus on examining the 5-12 $\mu$m region covered by JWST MIRI LRS for our examination of observable trends, because it is close to the peak of thermal emission of GJ~1214b and other planets in the temperature region where hazes are expected to be most dominant for sub-Neptunes. Furthermore, \citet{KemptonEtAl2023} estimated that 50-60\% of the thermal flux of this planet is emitted within the MIRI LRS bandpass.

\subsection{Phase curves}
\label{subsec:phasecurves}
\begin{figure*}
\begin{center}
\includegraphics[width=\textwidth]{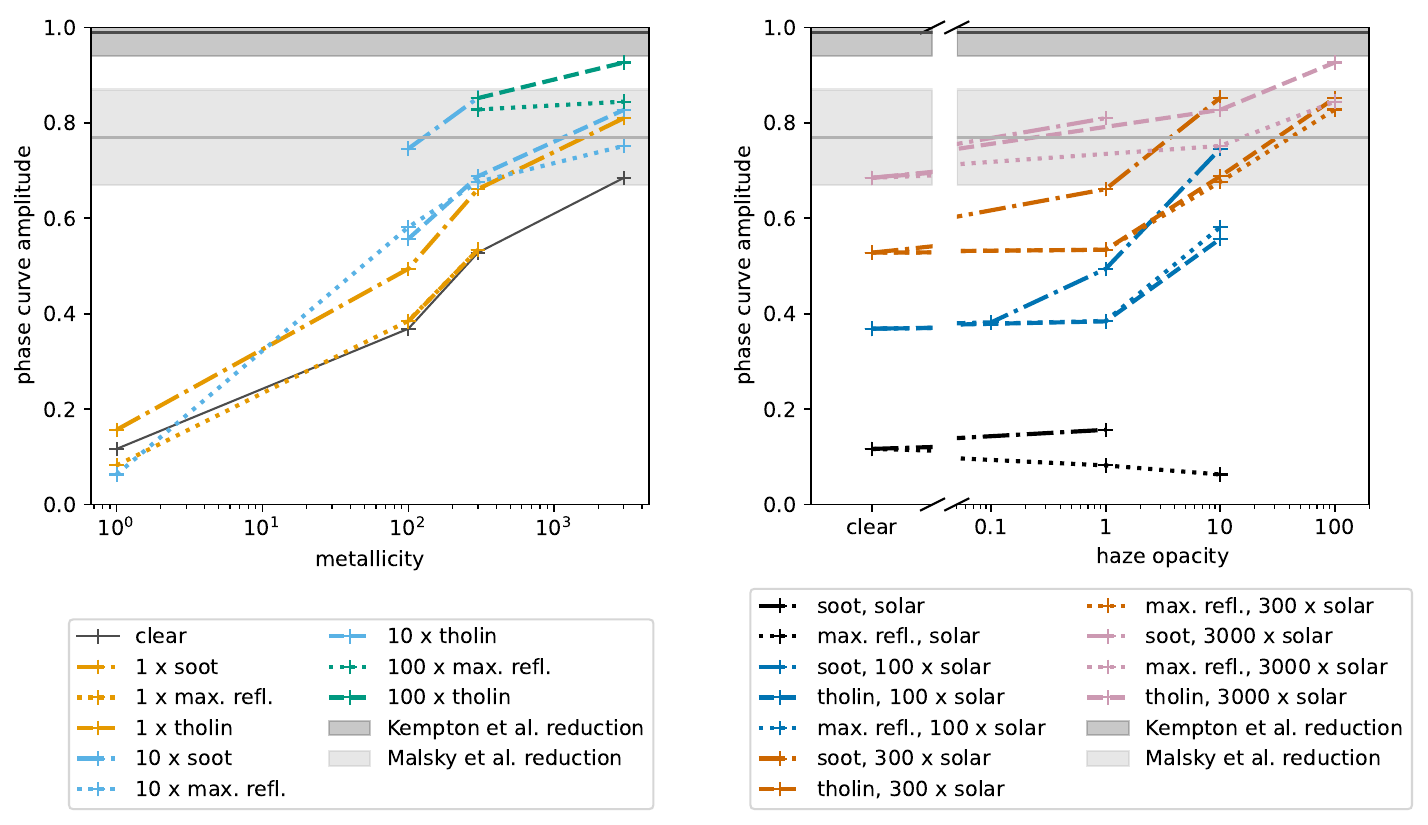}
\caption{White-light phase curve amplitudes in the JWST MIRI LRS bandpass of all simulations included in this study, plotted against metallicity (left panel) and haze opacity (right panel). The phase curve amplitudes generally increase with increasing metallicity and with increasing haze opacity. We define the phase curve amplitude as $(F_{\mathrm{max}}-F_\mathrm{min})/F_\mathrm{max}$, where $F_\mathrm{max}$ and $F_\mathrm{min}$ are the maximum and minimum values of the flux of the planet. The gray horizontal lines and shaded regions indicate the observed values from the Kempton et al. and Malsky et al. data reductions, including their 1-$\sigma$ uncertainty intervals.}
\label{fig:phasecurve_amplitudes}
\end{center}
\end{figure*}

\begin{figure*}
\begin{center}
\includegraphics[width=\textwidth]{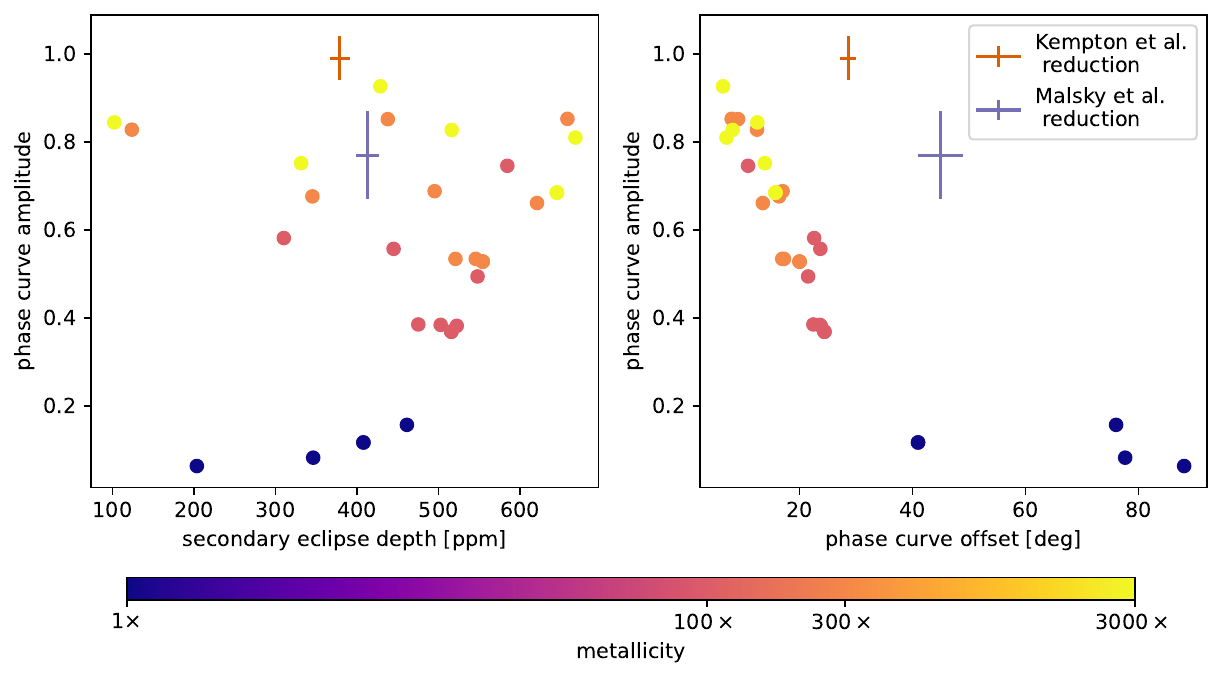}
\caption{White-light phase curve amplitudes in the JWST MIRI LRS bandpass of all simulations included in this study, plotted against secondary eclipse depth (left panel) and eastward phase curve offsets (right panel). Information on hazes has been removed to demonstrate that differences in haze opacity and haze optical properties lead to a spread in phase curve amplitudes, secondary eclipse depth and phase curve offset that make it impossible to determine metallicity from the white-light phase curve alone. The colorscale represents metallicity.}
\label{fig:amplitude_vs_eclipsedepth}
\end{center}
\end{figure*}

\begin{figure*}
\begin{center}
\includegraphics[width=\textwidth]{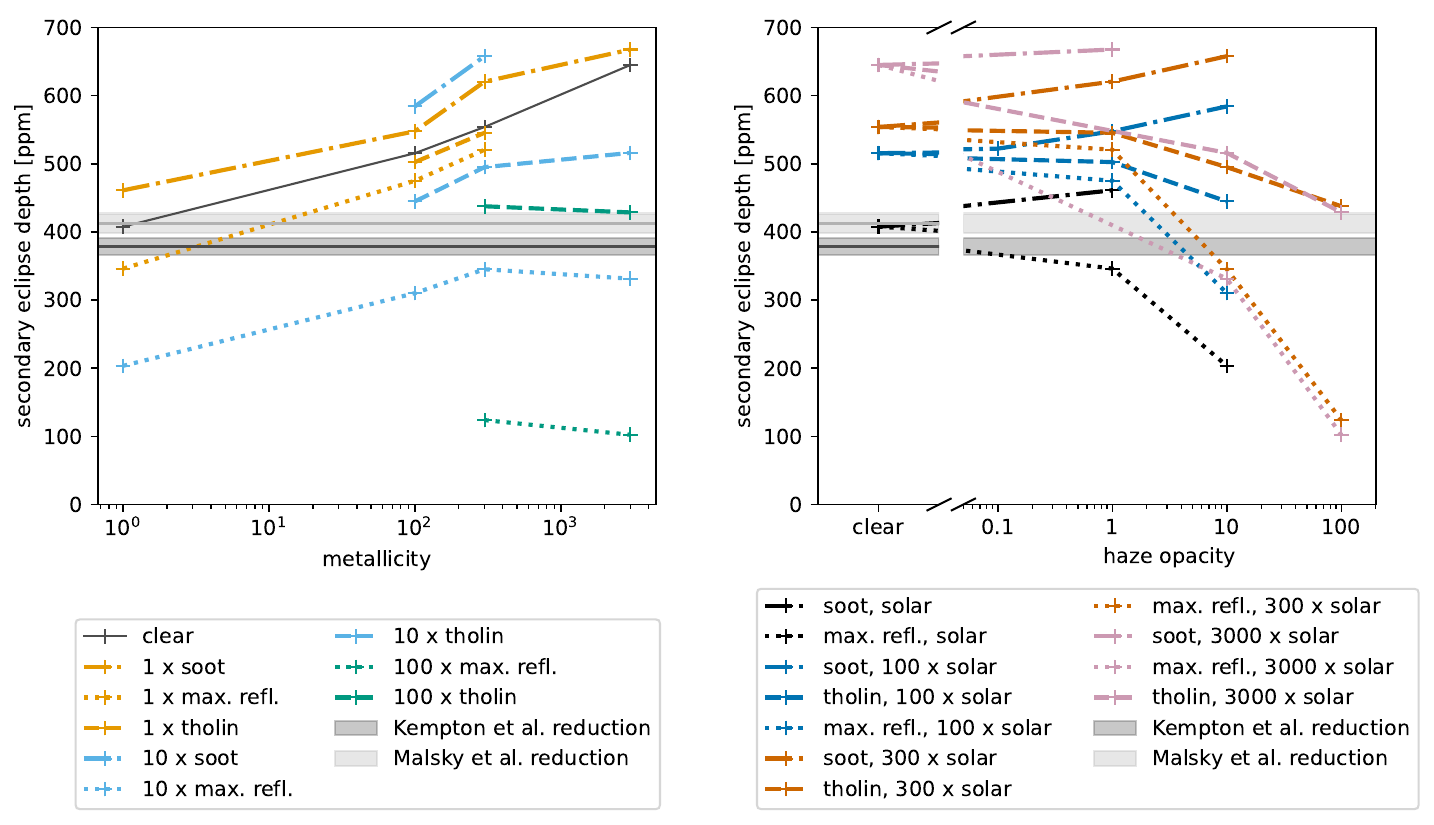}
\caption{White-light secondary eclipse depths in the JWST MIRI LRS bandpass of all simulations included in this study, plotted against metallicity (left panel) and haze opacity (right panel). Generally, absorbing hazes like soots increase the eclipse depth, while more scattering hazes like tholins and maximally reflective hazes decrease the eclipse depth. }
\label{fig:secondary_eclipse_depth}
\end{center}
\end{figure*}

\begin{figure*}
\begin{center}
\includegraphics[width=\textwidth]{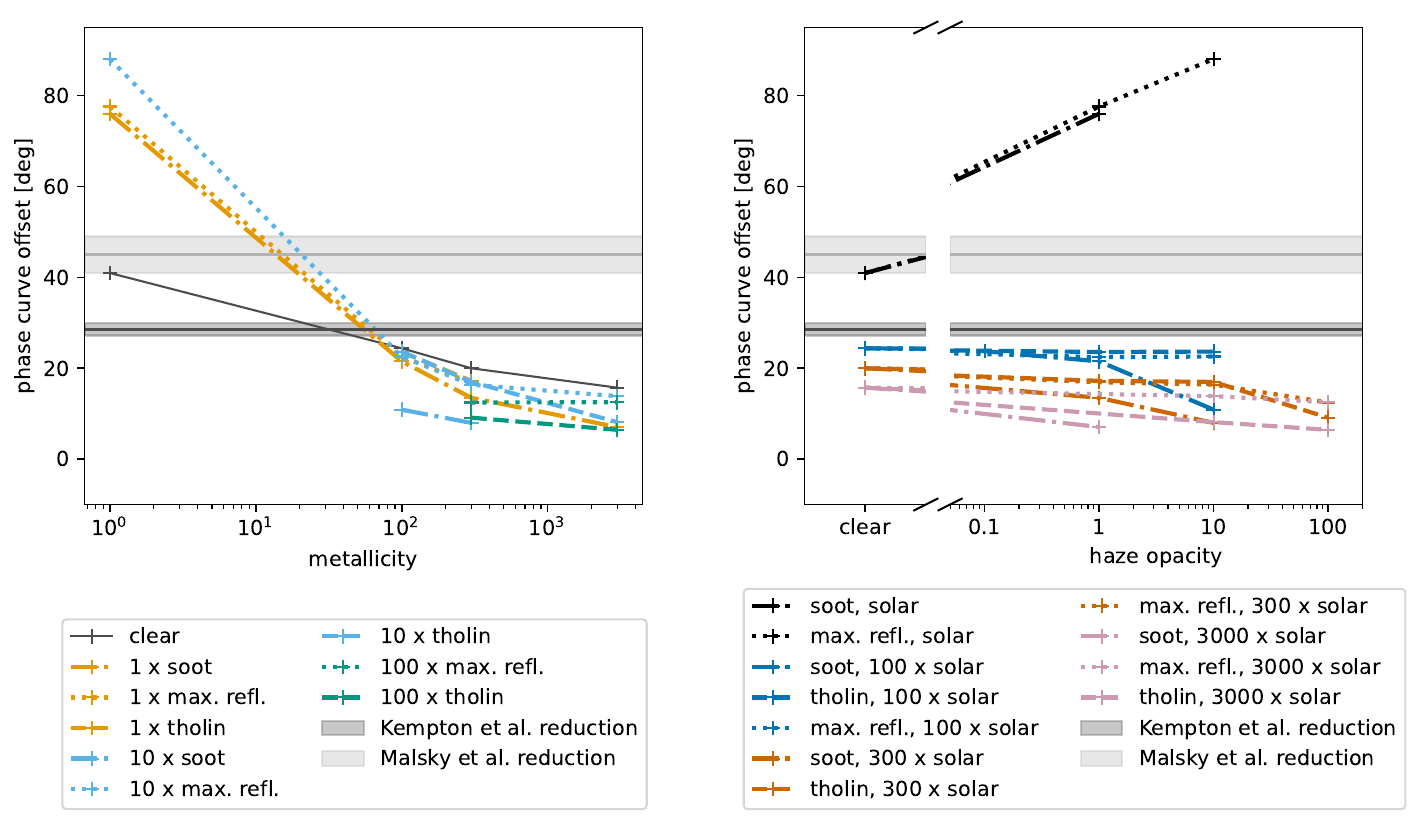}
\caption{Eastward white-light phase curve offsets  in the JWST MIRI LRS bandpass of all simulations included in this study, plotted against metallicity (left panel) and haze opacity (right panel).}
\label{fig:phasecurve_offset}
\end{center}
\end{figure*}
We start by focusing on properties that can be extracted directly from white-light phase curve observations. We first turn our attention to how photochemical hazes affect the relationship between phase curve amplitude and mean molecular weight  \citep[or, for our parametrization of atmospheric composition, equivalently, metallicity;][]{ZhangShowman2017}. The white-light phase curve amplitudes in the MIRI LRS bandpass are shown in Fig. \ref{fig:phasecurve_amplitudes}. We first point out that the clear-atmosphere simulations reproduce the expected trend of increasing phase curve amplitude with increasing metallicity. When hazes are added, the phase curve amplitude generally increases\footnote{The solar metallicity simulations with maximally reflective hazes are an exception to this: In this particular case, the phase curve amplitude drops ever so slightly.}. For any given combination of haze opacity and haze optical properties, the phase curve amplitude also increases monotonously with increasing metallicity. However, looking at the full set of simulations with varying haze opacities and haze properties, hazes add a considerable spread to the trend, with the difference between the clear-atmosphere amplitude and the amplitude of highest haze opacities with the same metallicity reaching up to 0.4. This spread is enough that it is no longer possible to uniquely determine the metallicity or mean molecular weight from the phase curve amplitude alone. However, the most extreme cases (solar vs. 3000$\times$solar) are still clearly separated.

When looking at the relationship between phase curve amplitude and haze opacity, it becomes obvious that for low and nominal haze opacities (0.1$\times$ and 1$\times$ the nominal haze opacity), the phase curve amplitude typically remains unchanged. When increasing the haze opacity enough to substantially affect the dayside thermal structure, the phase curve amplitude increases. For tholins and maximally reflective hazes, this transition happens between 1$\times$ and $10\times$ our nominal haze opacity. For soots, however, even the nominal haze opacity is enough to substantially increase the phase curve amplitude. This is consistent with Section \ref{subsec:temperaturestructure}, where we found that for soot, a thermal inversion already forms for the nominal haze opacity.

One might hope that the degeneracy between phase curve amplitude, haze opacity, and haze optical properties could be resolved by considering additional characteristics of the white-light phase curve such as secondary eclipse depth and the offset of the phase curve maximum with respect to secondary eclipse. However, these additional observables cannot fully break the degeneracy. This is demonstrated in Fig. \ref{fig:amplitude_vs_eclipsedepth}, which shows all simulations but information on haziness has been removed. There is a large scatter in secondary eclipse depth in addition to the scatter in amplitude. Some simulations with different metallicities are seen very close to each other in the amplitude-eclipse depth and amplitude-offset diagrams. Depending on the haze optical properties, the secondary eclipse depth can either increase for primarily absorbing hazes like soot or decrease for primarily reflecting hazes like tholins or our ``maximally reflective'' hazes (Fig \ref{fig:secondary_eclipse_depth}). In the case of reflecting hazes, it is thus possible to distinguish between the case of a clear atmosphere with higher metallicity and a hazy atmosphere with a lower metallicity with the same phase curve amplitude. Different metallicities with different levels of haze opacity, however, can still look very similar to each other. However, for soot hazes, the secondary eclipse depth increases with increasing haze opacity, as it would for a clear atmosphere with increasing metallicity. It is thus impossible to tell the difference between these two scenarios based on the white-light phase curve amplitude and secondary eclipse depth alone.

Further, we note that the phase curve offset (Fig. \ref{fig:phasecurve_offset}) is unlikely to provide useful information on metallicity or haze opacities in most cases. The phase curve offset is strongly correlated with the phase curve amplitude (Fig. \ref{fig:amplitude_vs_eclipsedepth}, right panel) and thus does not provide any complementary information to the phase curve amplitude. In addition the offset is already small ($\approx 25^{\circ}$) in the 100$\times$~solar clear-atmosphere simulation. While it further decreases to $\approx 15^{\circ}$ at 3000$\times$~solar (clear-atmosphere), the difference of 10$^{\circ}$ between these two models is comparable to typical observational uncertainties of the phase curve offset. For example, the $1 \sigma$ uncertainty on the GJ~1214b phase curve in the Malsky et al. reduction is $\pm4^{\circ}$. Furthermore, observational publications often underestimate the uncertainty on phase curve offsets because they do not take pipeline-induced uncertainties into account \citep{BellEtAl2021SpitzerReanalysis}. Thus, even in the clear-atmosphere case, it is already hard to distinguish between different metallicities based on the phase curve offset. The additional spread introduced by the different levels haze opacities further ruins the already-slim prospects of determining metallicity based on the phase offset.

All solar metallicity simulations, clear and hazy, exhibit a much larger eastward phase curve offset than all other simulations. All of the solar-metallicity hazy models further can be clearly distinguished from the clear-atmosphere case by having an even larger phase curve offset. The larger phase curve offset is driven by a stronger eastward jet. In the case of the solar metallicity, the jet speed is very sensitive to even the nominal haze opacity and the resulting increase in the hotspot offset is larger than the effect of probing higher layers (with a smaller hotspot offset) due to the added haze opacity. However, we caution that all of these models also exhibit a very low phase curve amplitude. For low-amplitude phase curves, the offset generally is harder to measure observationally and the observational uncertainty will be larger. Past studies have also shown the phase curve offset can be strongly affected by inhomogeneous cloud coverage \citep{ParmentierEtAl2021}, an effect not included in our models.

We thus conclude that a band-integrated phase curve alone may be insufficient to determine the metallicity or mean molecular weight of the atmosphere in the presence of hazes. Next, we turn to emission spectra. We will further discuss potential observational strategies for determining atmospheric composition for hazy atmospheres in Section \ref{subsec:observingstrategies}.

\subsection{Emission spectra}
\label{subsec:emissionspectra}
\begin{figure}
\begin{center}
\includegraphics[width=\columnwidth]{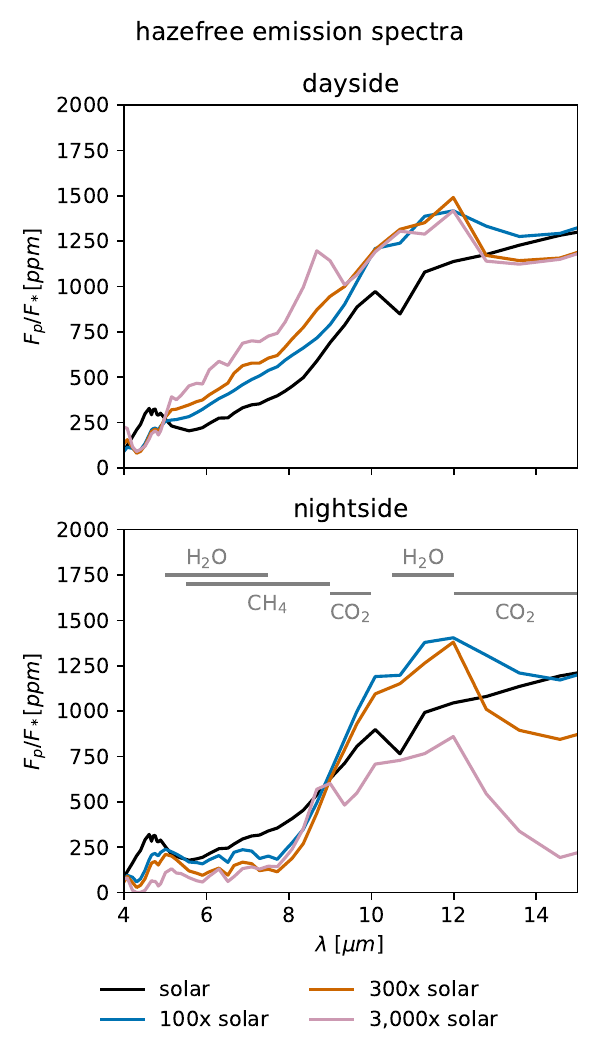}
\caption{Dayside and nightside emission spectra of the hazefree simulations.}
\label{fig:spectra_hazefree}
\end{center}
\end{figure}
Before we examine the effect of photochemical hazes on dayside and nightside spectra, we provide a brief overview of the  clear-atmosphere emission spectra (Fig. \ref{fig:spectra_hazefree}). Contribution functions for the clear-atmosphere cases are shown in the appendix in Fig. \ref{fig:contributionfunctions}. Water and methane are responsible for the most prominent absorption feature visible at all metallicities, spanning from approximately 5 to 9 $\mu$m. Water here dominates the spectrum below $\approx7 \mu$m and methane above $\approx 7 \mu$m, with both absorption bands smoothly blending into each other. With increasing metallicity, CO$_2$ gains increasing importance \citep[see e.g., Fig. 2 in][]{GaoEtAl2023GJ1214b}, first by increasing absorption in the 15 $\mu$m band (which almost, but not quite, reaches into the MIRI LRS wavelength range), then by also adding an additional absorption feature between 9 and 10 $\mu$m in the 3,000$\times$ solar spectrum.

With increasing metallicity, the dayside emission increases at almost all wavelengths, while the change in the nightside spectra is more complicated. On the nightside, the emission drops dramatically in the H$_2$O and CH$_4$ band as metallicity increases. However, in the spectral window beyond 9 $\mu$m, the emission first increases substantially for the 100$\times$ and 300$\times$ solar simulations, with an emitted flux close to the dayside flux in this spectral region. These spectral windows are probing deeper layers where atmospheric circulation is efficient at transporting heat and day-night gradients are small. Only with the highly increased CO$_2$ abundance in the 3,000$\times$~solar simulation does the nightside flux drop off in this wavelength region, falling below the flux of the solar metallicity simulation.

\subsubsection{Dayside spectra}
\begin{figure*}
\begin{center}
\includegraphics[width=\textwidth]{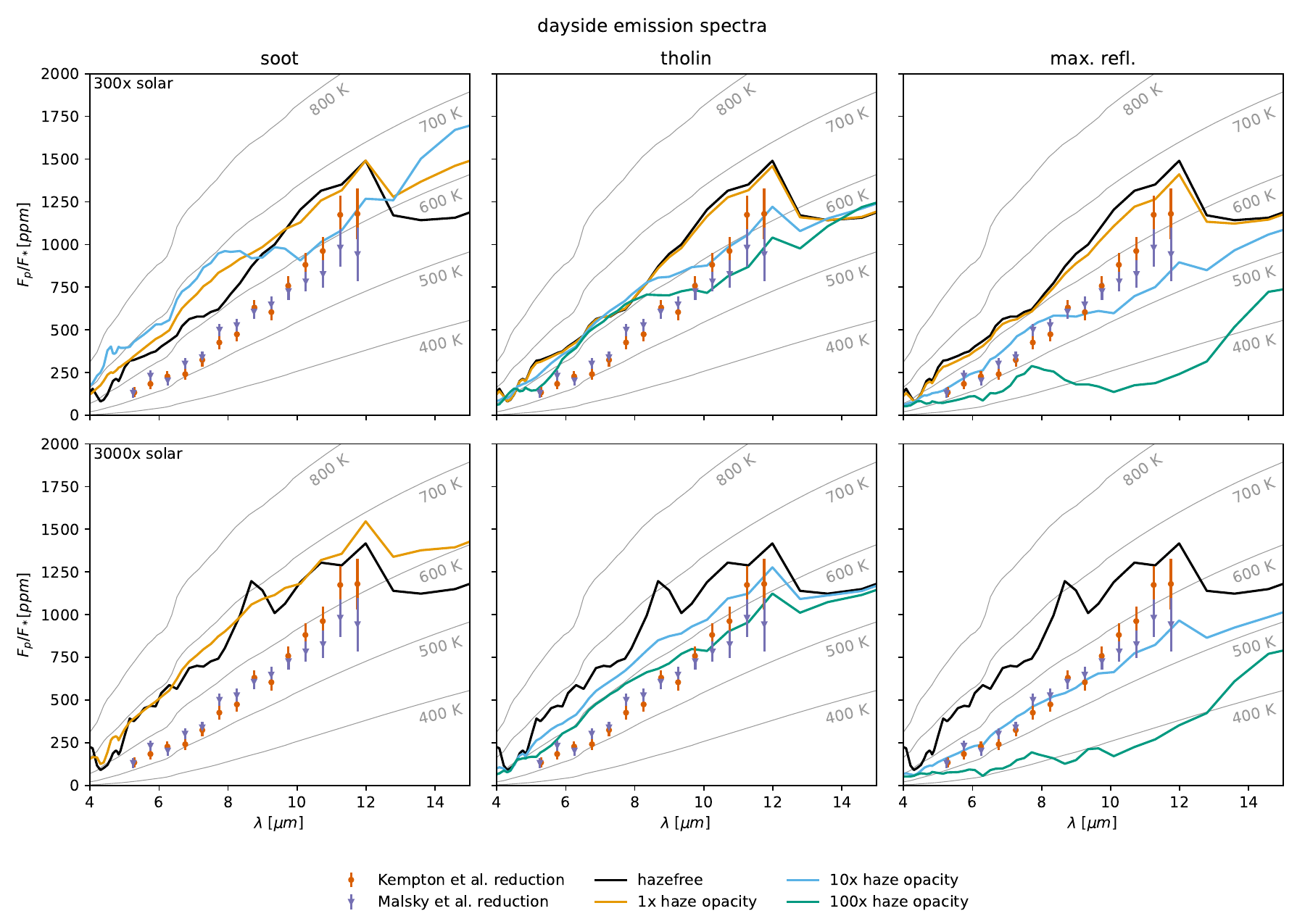}
\caption{Dayside emission spectra at secondary eclipse for multiple haze opacities and haze properties, for a metallicity of 300$\times$~solar (top row) and 3000$\times$~solar (bottom row). With increasing haze opacity, spectral features seen as absorption features in the haze-free case gradually turn into emission features. Gray lines indicate spectra for a planet emitting blackbody radiation at multiple temperatures (indicated by the labels) for comparison. The JWST MIRI LRS secondary eclipse spectrum of \citet{KemptonEtAl2023} and the alternative data reduction of \MalskyEtAlT are shown as red circles and purple triangles, respectively, with errorbars. }
\label{fig:spectra_day_300xsolar}
\end{center}
\end{figure*}
Even though the change in the band-integrated emitted flux due to hazes varies dramatically between the three sets of haze optical properties we examined, we identify a common pattern in how the spectra change with increasing haze opacity for all three haze properties. First, the amplitude of absorption features decreases until the spectrum is almost featureless (i.e., probing an isothermal portion of the atmosphere). Then, emission features appear (Fig. \ref{fig:spectra_day_300xsolar}). This behavior is consistent with previous studies on the effect of photochemical hazes on emission spectra of sub-Neptunes in 1D \citep{MorleyEtAl2015} and hot Jupiters in 1D \citep{LavvasArfaux2021,ArfauxLavvas2022} and 3D \citep{SteinrueckEtAl2023}. However, the overall spectral shape and where this transition happens depend on haze properties and metallicity.

For soot hazes, a thermal inversion shows up at lower haze opacities---even at the nominal haze opacity---compared to the less absorbing hazes, resulting in emission features. The only exception to this is the 3000x solar model, which is almost completely featureless from 5 to 10 microns and exhibits small absorption features at shorter and longer wavelengths. Because the thermal inversion is driven by additional heating through absorption by soot hazes (see Section \ref{subsec:temperaturestructure}), the emitted planetary flux increases substantially in molecular absorption bands such as the water and methane bands from around 6 to 8 $\mu$m.

For the maximally reflective hazes, as described in Section \ref{subsec:temperaturestructure}, the thermal inversion is driven entirely by cooling in deeper layers due to incoming starlight being reflected at high altitudes. Thus, the overall dayside emission drops off dramatically with increasing haze opacity. At the nominal haze opacity, there is a decreased amplitude of absorption features at all metallicities, but no emission features are visible yet. For 10x the nominal haze opacity, the water/methane feature from 5-9 $\mu$m shows up in emission for all metallicities except for the 3000$\times$ solar case. Increasing the haze opacity even further to 100x leads to a drastic decrease in the overall flux and even stronger emission features. For this haze opacity, the 3000$\times$ solar case also exhibits emission features.

We note that because the emissivity of the maximally reflective hazes effectively is zero due to their high single scattering albedo, no photons are emitted from the haze particles \citep[e.g.,][]{TaylorParmentier2021ScatteringInEmission}. Consequently, the emission effectively emerges from regions with significant gas-phase opacity, and thus from much deeper regions of the atmosphere than in the soot case. Therefore, even though the optical depth (which includes scattering and absorption) reaches unity at $\approx 10^{-5}$~bar in the 100$\times$ haze opacity case, the emission features indeed probe the thermal inversion at $\approx 10^{-3}$ to $10^{-4}$~bar in the $300\times$~solar and $3000\times$~solar metallicity cases.

For tholin hazes, the dayside spectrum transitions from absorption to emission features at similar haze opacities as for the maximally reflective hazes. However, due to the additional heating due to absorption by hazes at low pressures, the emitted flux does not drop nearly as dramatically with increasing haze opacity as in the maximally reflective case. In fact, in the center of the water and methane bands, the emitted flux stays roughly the same or slightly increases, while it drops by less to ~30\% outside the band even in the 100$\times$ haze production cases.
As with the other haze properties, for the 3000x solar simulation, a higher haze opacity (100x) is needed to produce emission features, again consistent with the fact that at haze opacities $<100\times$ the nominal one, no thermal inversion forms. This again confirms the picture emerging in previous sections that for higher metallicities, a larger haze opacity is needed to alter the atmospheric structure than for lower metallicities.

\subsubsection{Nightside spectra}
\begin{figure}
\begin{center}
\includegraphics[width=\columnwidth]{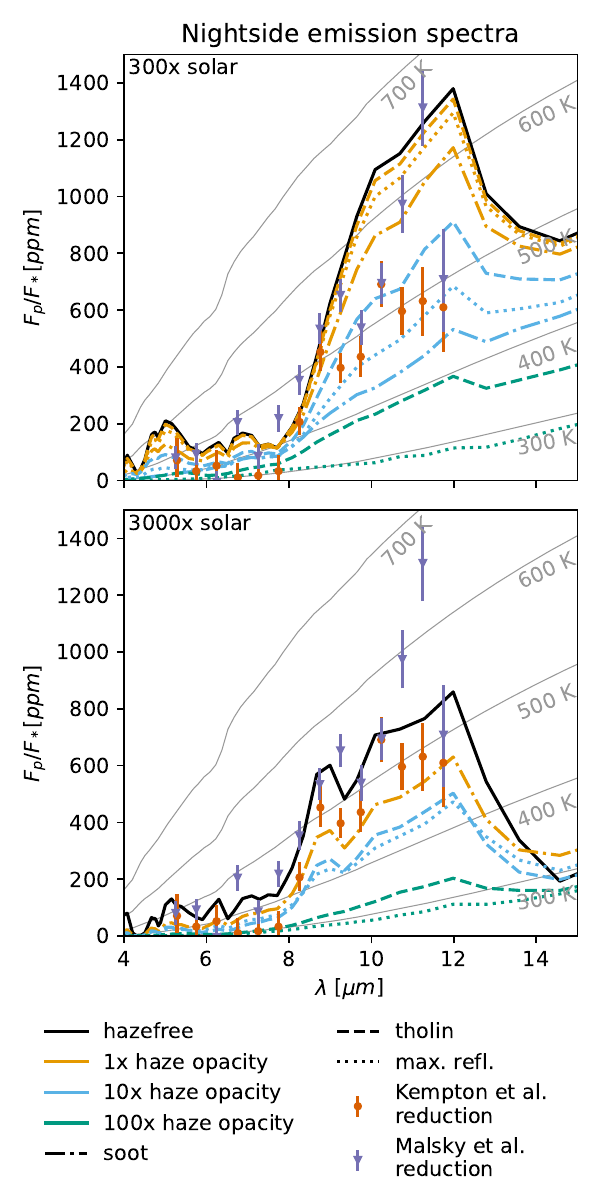}
\caption{Nightside emission spectra for multiple haze opacities and haze properties, for a metallicity of 300$\times$~solar (top) and 3000$\times$~solar (bottom). With increasing haze opacity, the nightside flux drops off due to a combination of cooling of the deep atmosphere and emission emerging from higher altitudes. Gray lines indicate spectra for a planet emitting blackbody radiation at multiple temperatures (indicated by the labels) for comparison. The JWST MIRI LRS nightside spectrum of \citet{KemptonEtAl2023} and the alternative data reduction of \MalskyEtAlT are shown as red circles and purple triangles, respectively, with errorbars. }
\label{fig:spectra_night}
\end{center}
\end{figure}
The spectral shape of the nightside emission spectra (Fig. \ref{fig:spectra_night}) remains similar to the clear-atmosphere spectrum except for very high haze opacities. However, the emitted flux is reduced for all hazy simulations, with a larger reduction in emitted flux for higher haze opacities. Two factors contribute to this reduction in flux: the cooling of deep layers of the atmosphere and the additional haze opacity resulting in the emitted radiation emerging from lower pressure levels with lower temperatures.

When comparing identical haze opacities, the reduction in flux is smallest for tholins and largest for soots. Tholins have a lower extinction cross section than soots at most wavelengths, thus the outgoing emission probes deeper, hotter layers than for soot or maximally reflective hazes. Because we assumed an extinction cross section identical to soot for the maximally reflective hazes, the outgoing emission should emerge from roughly similar regions in the atmosphere for soot and maximally reflective hazes. The lower nightside flux in the soot case thus can be attributed to the stronger cooling of the deep atmosphere. As the haze opacity increases, the region probed by the outgoing emission shifts to higher regions in the atmosphere, at some point reaching the region in which the temperature profile is close to isothermal (Fig. \ref{fig:temperatureprofiles_hazy}). This explains why the emission spectra are almost featureless for the 100$\times$ haze production cases.

\section{Comparison to GJ~1214b phase curve observations}
\label{sec:gj1214b_comparison}

\begin{figure}
\begin{center}
\plotone{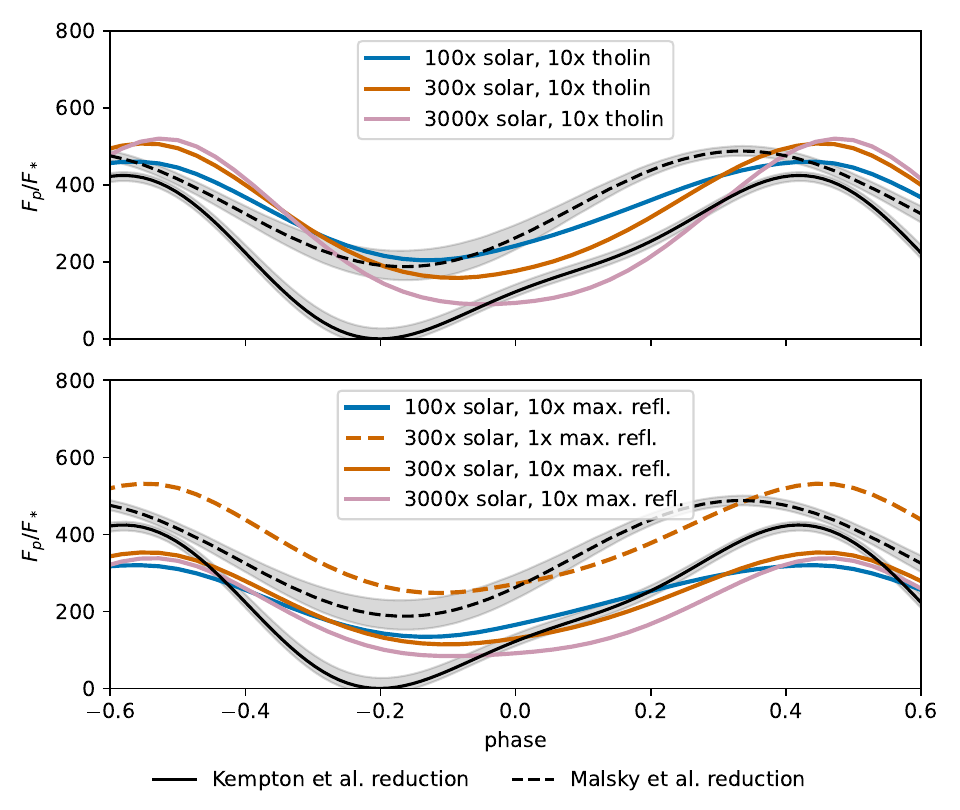}
\caption{Models which provide a relatively good match to the observed JWST MIRI LRS white-light phase curve of GJ~1214b, compared to the data reductions of \citet{KemptonEtAl2023} and \MalskyEtAlT. Generally, models with a metallicity $\geq100\times$~solar and tholins or maximally reflective hazes provide the closest qualitative match. }
\label{fig:whitelight_bestmatch}
\end{center}
\end{figure}
In this section, we compare our models to the recent phase curve observations of GJ~1214b with JWST MIRI/LRS (Fig. \ref{fig:whitelight_bestmatch}). A comparison between most of our models and the observed phase curve of GJ~1214b was already presented in \citet{KemptonEtAl2023}. They conclude that both a high metallicity ($\gtrapprox 100\times$~solar) and a high  ($10\times$~nominal) haze opacity with hazes that are more reflective than tholins are needed to qualitatively match the white-light phase curve. These models roughly reproduce the observed dayside flux and phase curve offset, with observations falling in between the tholin and ``maximally reflective'' models, and also produce a large phase curve amplitude. None of the models, however, predict nightside fluxes as low as in the data reduction presented in that paper. It is worth noting that \citet{KemptonEtAl2023} stressed that the nightside flux was the most uncertain aspect of the observational analysis.

One aspect not discussed in \citet{KemptonEtAl2023} is that the dayside spectra of almost all models that qualitatively match the phase curve (i.e., that have a haze opacity high enough to cool down the dayside flux substantially) exhibit quite prominent emission features. The only exception to that is the $3000\times$~solar model with $10\times$~haze opacity, which shows an almost-featureless spectrum with very subtle emission features. In contrast, the secondary eclipse spectrum reported in \citet{KemptonEtAl2023} shows subtle absorption features. The observed shape of the nightside spectrum matches models with a metallicity $\gtrapprox100\times$~solar and $\approx 10\times$ the nominal haze opacity qualitatively, though the flux between 6 to 8 $\mu$m dips noticeably below the values predicted by these models.

\MalskyEtAlT present an alternative data reduction that includes a previously unidentified potential systematic (a sudden drop in flux near the phase curve minimum). Including this additional systematic leads to a much higher nightside flux. The phase curve amplitude and flux at the phase curve maximum are now close to a range of models, especially the 100$\times$ and 300$\times$ solar models with tholins and 10$\times$ the nominal haze opacity. We note that in terms of amplitude and maximum flux, the phase curve also lies between the 300$\times$~solar models with maximally reflective hazes with 1$\times$ and $10\times$ the nominal haze opacity and is relatively close to the 3000$\times$~solar model with tholins and $10\times$ the nominal haze production. However, all of the models predict a phase curve offset that is much smaller than the observed offset using the \MalskyEtAlT reanalysis.

The shape of the dayside spectrum also changes with the reanalysis, becoming closer to a blackbody spectrum (Fig. \ref{fig:spectra_day_300xsolar}). This could be indicative of hazes on the dayside. However, the tension between our models predicting substantial emission features and the lack of emission features in the observed spectrum remains. 
The shape of the nightside spectrum matches the models that provide the closest match to the white-light phase curve relatively well when ignoring the likely unreliable data points with $\lambda>10.5$~$\mu$m. Past 10.5~$\mu$m, the data exhibited considerable red noise and the phase curve fits in these wavelength bins were poor (see \MalskyEtAlT).

\begin{figure*}
\begin{center}
\includegraphics[width=\textwidth]{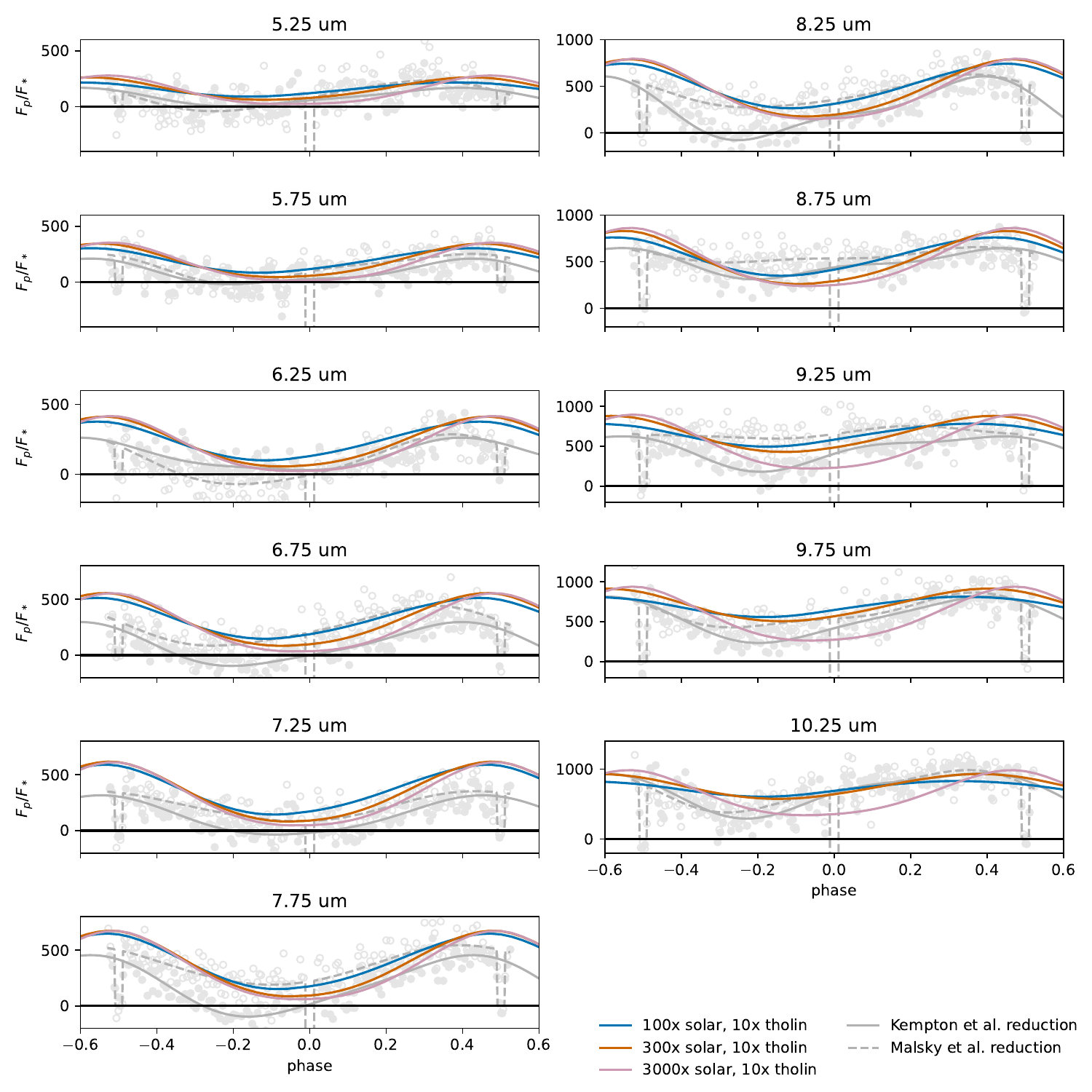}
\caption{Spectral phase curves for a selection of models with tholin hazes that provide a relatively good match to the observed white-light phase curve. The models substantially overpredict the emitted flux for a large fraction of the phase curve in spectral regions that show emission features in the secondary eclipse spectrum (6-8.5 $\mu$m). The observed spectral phasecurves are plotted in the background in light gray with filled circles and solid lines representing the \citet{KemptonEtAl2023} reduction and empty circles and dashed lines the \MalskyEtAlT reduction. We note that due to the relatively low resolution of our post-processing, the spectral bins of the models do not line up exactly with the spectral bins of the observations (see Appendix for more detail).}
\label{fig:spectralphasecurve_tholin}
\end{center}
\end{figure*}

\begin{figure*}
\begin{center}
\includegraphics[width=\textwidth]{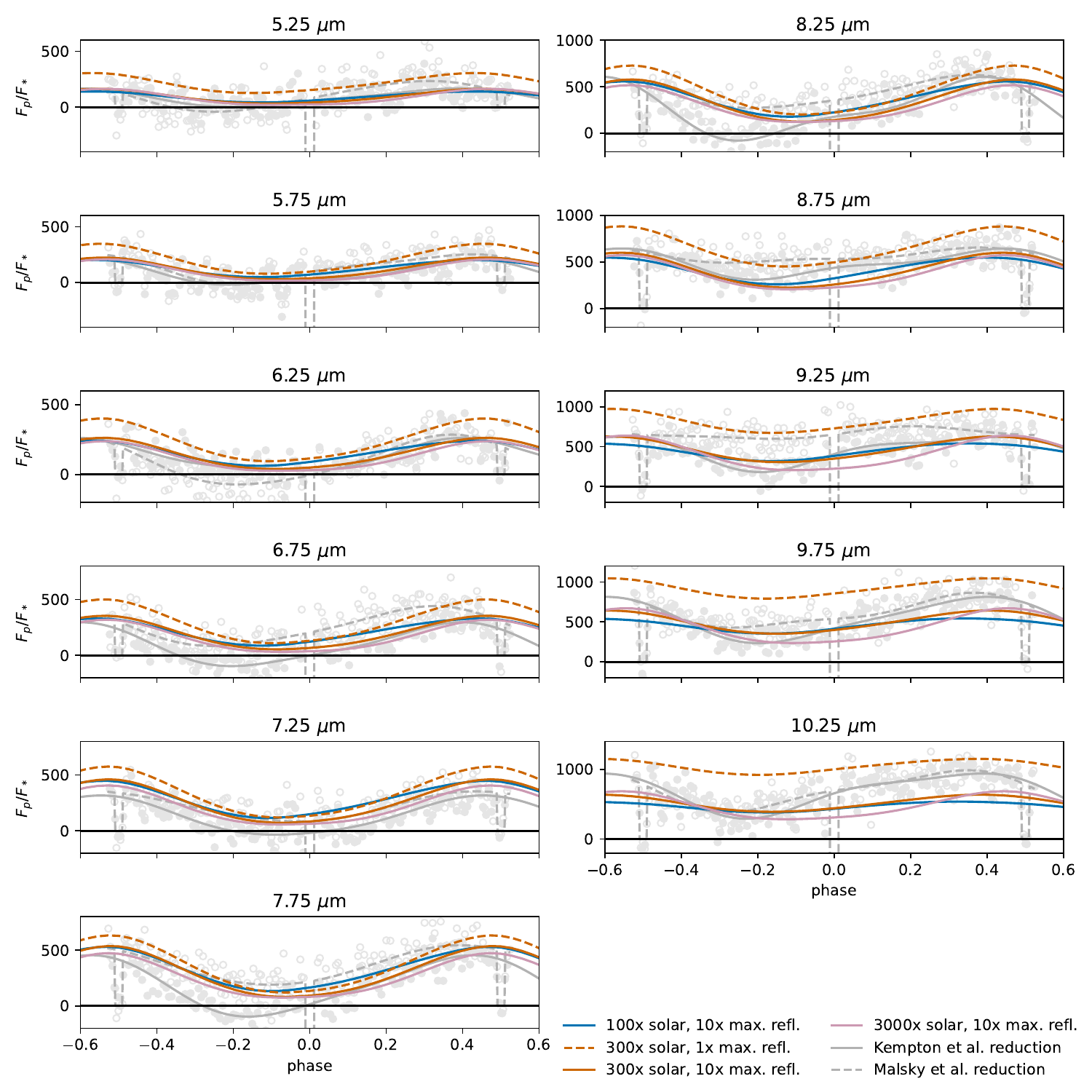}
\caption{As in Fig. \ref{fig:spectralphasecurve_tholin} but for a selection of models with maximally reflective hazes that provide a relatively good match to the observed white-light phase curve. The models provide a good qualitative match at wavelengths $<$9.5~$\mu$m (Kempton reduction)/$<$9.5~$\mu$m (Malsky reduction) but underestimate the flux, and in some channels, amplitude in the longest wavelength channels. }
\label{fig:spectralphasecurve_maxrefl}
\end{center}
\end{figure*}

To utilize the spectral information from other phases as well, we proceed to look at the spectral phase curves of the models we identified as qualitatively matching the white-light phase curve best (Fig. \ref{fig:spectralphasecurve_tholin} and \ref{fig:spectralphasecurve_maxrefl}). For the simulations with tholins among this set (Fig. \ref{fig:spectralphasecurve_tholin}), generally, the models over-predict the emitted flux at wavelengths $<8.5$~$\mu$m for a substantial portion of the phase curve (centered at the dayside). At wavelengths $>9.5$~$\mu$m, in turn, the 3000$\times$ solar phase curve matches the observed maximum and minimum fluxes (especially the Kempton et al. reduction) well but is shifted towards a smaller phase offset than observed, while the lower metallicity simulations exhibit a lower phase curve amplitude and higher nightside fluxes than observed. The pattern of over-predicting flux within the water and methane absorption bands while matching the phase curve better in channels outside these absorption bands can be attributed to the emission features exhibited by the models but not seen in the data. Seeing this discrepancy with observations arise in the spectral phase curve as well demonstrates that the thermal inversion dominates the model spectra throughout the entire dayside. Thus, the mismatch to the observed spectral shape is not just limited to a narrow phase region near the secondary eclipse.

The simulations with maximally reflective hazes that provide a reasonable match to the white-light phase curve, in turn, exhibit a closer match to the observed spectral phase curve at wavelengths $<9$~$\mu$m but  under-predict the flux at almost all phases at longer wavelengths. For the 100$\times$ and 300$\times$~solar simulations, the phase curve amplitude also is smaller than observed at these wavelengths, while the 3000$\times$~solar phase curve provides a somewhat better match to observations in the longest wavelength channels. The pattern of having lower-than-observed flux at wavelengths $<9$~$\mu$m can also be attributed to the presence of emission features in most model-predicted spectra, which is not seen in the observations. An exception to this behavior is the 300$\times$~solar simulation with nominal haze opacity, which over-predicts the flux in all channels at all phases except for the 7.75~$\mu$m and 8.25~$\mu$m channels, in which the emitted flux falls below the Malsky et al. reduction but above the Kempton et al. reduction for the phases in which primarily the nightside is visible.

We further note that the trend of higher phase curve amplitudes with higher metallicities, when the haze properties are held constant, hold true even in the narrow wavelength channels.

To summarize, some of our models are able to match multiple aspects of the observed phase curve of GJ~1214b well, especially the low dayside flux, the shape of the nightside spectrum, and, in the case of the \MalskyEtAlT reanalysis, the phase curve amplitude. However, remaining tensions are the observed lack of emission features on the dayside, the low nightside flux in the case of the Kempton data reduction and the high observed phase curve offset in the case of the Malsky reanalysis. One simulation (3000$\times$~solar, maximally reflective hazes, 10$\times$~haze opacity) only shows very subtle emission features and thus qualitatively provides a somewhat better match to the secondary eclipse spectrum and spectral phase curve. Yet, this model overall overestimates the albedo of the planet and thus emits too little flux at almost all phases when integrated across the MIRI bandpass. Our model may be too simplistic to fully match the observations. For example, simultaneously including condensate clouds could lower the haze production rate that is required to explain the low dayside flux \MalskyEtAlP and result in a more blackbody-like spectrum. Using a more consistent haze profile derived from microphysics models run for each atmospheric composition and haze production rate (rather than applying the same haze profile to all atmospheric compositions and scaling the haze optical depth) could also change the vertical temperature structure and thus the strength of the thermal inversion. Spatially inhomogeneous hazes may also be possible \citep{SteinrueckEtAl2021}, and including a dynamic feedback between atmospheric transport of hazes, atmospheric circulation, and temperature structure could further impact the emission spectra and phase curve \citep{SteinrueckEtAl2023}. Finally, we note that the haze optical properties remain highly uncertain. While we explored multiple limiting cases, exploring a range of possible optical properties, such as those presented in recent studies \citep{HeEtAl2024WaterRichOpticalProperties,CorralesGavilanEtAl2023} may also improve the match to observations.

\section{Recommended observing strategy}
\label{subsec:observingstrategies}
\begin{figure*}
\begin{center}
\includegraphics[width=\textwidth]{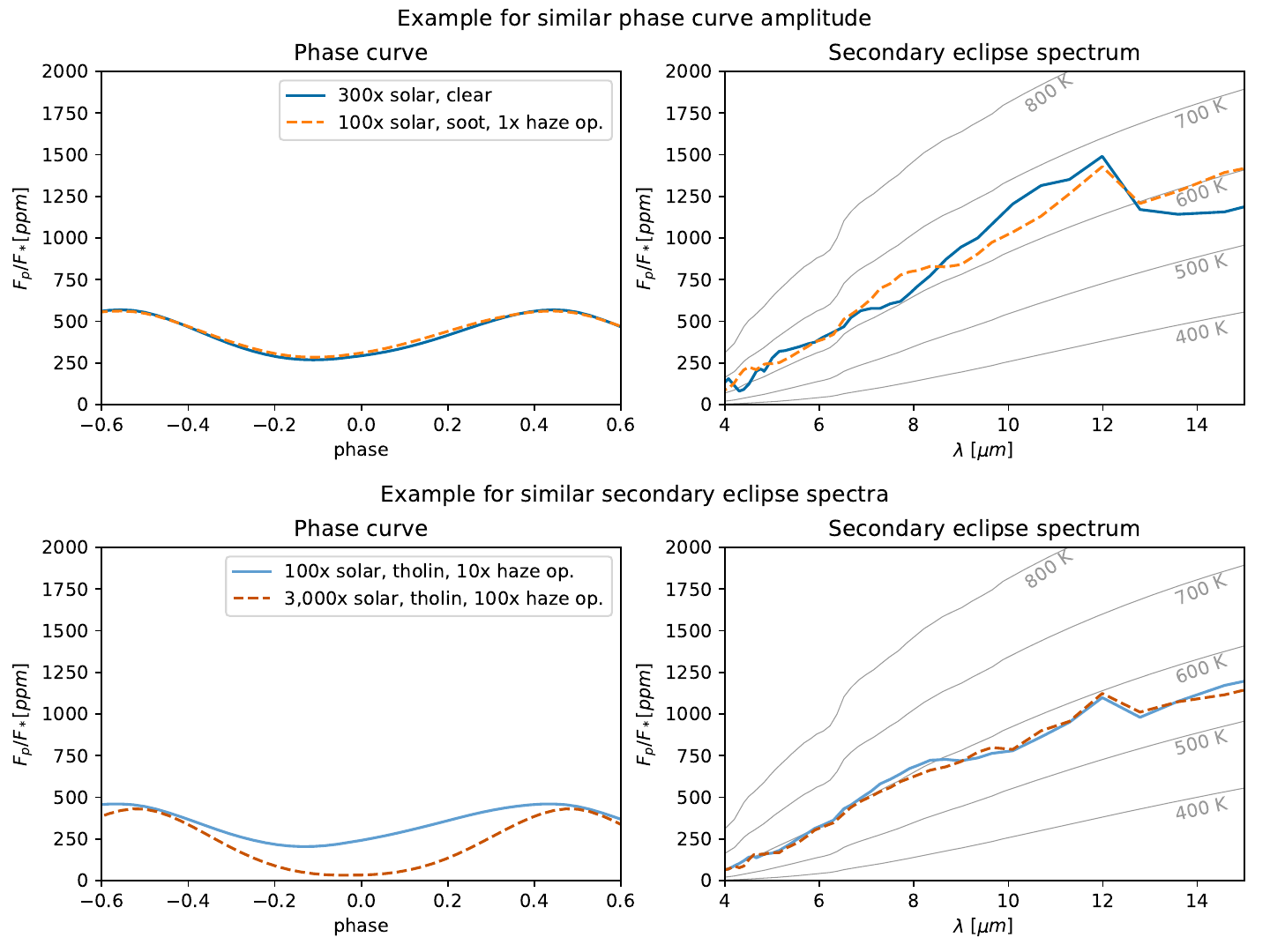}
\caption{This figure demonstrates the advantages of our recommended observing strategy of combining a white-light phase curve with a secondary eclipse spectrum. The top row shows an example of two models with different atmospheric composition with nearly identical phase curves but differing emission spectra, while the bottom row features an example of two models with nearly identical emission spectra but vastly differing white-light phase curves. The left column each shows the white-light phase curve in the MIRI bandpass and the right column the secondary eclipse spectrum.
}
\label{fig:observing_strategy}
\end{center}
\end{figure*}
As discussed in Section \ref{subsec:phasecurves}, photochemical hazes add a considerable spread to the phase curve amplitude--mean molecular weight/phase curve amplitude--metallicity trend, making it impossible to distinguish between a high-metallicity clear atmosphere and a lower-metallicity atmosphere with soot hazes. At the same time, intermediate\footnote{We use the term ``intermediate'' to mean intermediate within our chosen grid of parameters. However, when comparing to the predicted photochemical mass flux of haze precursors in photochemistry simulations, assuming haze yield rates similar to the Solar System, and assuming that the haze opacity roughly scales with haze production rate, the nominal (1$\times$~haze opacity) scenario would be considered a typical assumption and the 10$\times$~haze opacity scenario would be considered high, while the 100$\times$~haze opacity scenario would be considered extremely high--comparable to the surprisingly high haze production rates inferred for GJ~1214b by \citep{GaoEtAl2023GJ1214b} and \citep{OhnoEtAl2024GJ1214b}.} haze opacities can lead to an isothermal atmosphere on the dayside of the planet in pressure regions probed by emission spectroscopy, and thus a blackbody-like emission spectrum, again prohibiting the determination of the composition of the atmosphere. Here, we argue that a combination of a band-integrated phase curve and a secondary eclipse spectrum can break the degeneracy between metallicity and haze opacity and thus is a promising observing strategy for sub-Neptunes.

In Section \ref{subsec:phasecurves}, we observed that generally, the phase curve amplitude will only be affected by hazes if the haze opacity is large enough to affect the temperature structure of the atmosphere. In Section \ref{subsec:emissionspectra}, we found that for these cases, the dayside emission spectrum shows dramatically reduced absorption features, an almost featureless spectrum, or emission features. Thus, the dayside spectrum can inform the interpretation of the phase curve: If there are prominent absorption features present, with an amplitude comparable to what is expected for a clear atmosphere, it is safe to use the clear-atmosphere phase curve amplitude--metallicity relationship to deduce the metallicity. However, if emission features are present or the dayside spectrum looks blackbody-like, the phase curve should be compared to GCM simulations that include photochemical hazes for interpretation. An example, comparing the emission spectra of two atmospheres with nearly identical phase curves, is shown in the upper row of Fig. \ref{fig:observing_strategy}. The emission spectrum of the higher-metallicity, clear-atmosphere model exhibits easily visible absorption features, while for the hazy, lower-metallicity atmosphere, the same spectral features show up in emission.

Vice versa, a white-light phase curve adds complementary information to the secondary eclipse spectrum and can be crucial for narrowing down the atmospheric composition. As we demonstrated in Section \ref{subsec:emissionspectra}, intermediate (10$\times$) haze opacities can lead to almost featureless emission spectra that contain little information on the atmospheric composition. In these cases, a phase curve can distinguish between a higher-metallicity and a lower-metallicity scenario. An example is shown in the bottom row of Fig. \ref{fig:observing_strategy}.

For high haze opacities, our models predict strong emission features. In this scenario, it should be possible to obtain constraints on atmospheric composition and haze production rate from the emission features in the secondary eclipse spectrum. However, a caveat to this is that our models seem to overpredict the strength of emission features when compared to the emission spectrum of GJ~1214b. A more thorough examination of the scenario of high haze production rates with more complex haze models (e.g., considering the change in particle size distribution, and thus, opacity structure, with changing metallicity and haze production rate), may thus be necessary to correctly interpret scenarios with hints of high haze opacity.

Overall, we conclude that the combination of a white-light phase curve and a secondary eclipse spectrum is promising for characterizing the atmospheres of sub-Neptunes. 
For some sub-Neptunes, like GJ~1214b, the signal-to-noise is good enough that a meaningful dayside spectrum can be obtained from a single secondary eclipse or two secondary eclipses. For these planets, a  secondary eclipse spectrum will automatically be included with a phase curve observation.\footnote{For the very few targets that are bright enough that a spectral phase curve can be obtained, the spectral phase curve and in particular the nightside spectrum can be highly valuable in determining the composition of the atmosphere as well. Nightside spectra often will still exhibit spectral features even in the case of a featureless secondary eclipse spectrum, as evidenced both by our models (Fig. \ref{fig:spectra_night} and the observations of GJ~1214b. However, even for the best-case scenario of GJ~1214b, the nightside spectrum and spectral phase curves have significant uncertainty. The number of sub-Neptunes for which a spectral phase curve is feasible thus is severely limited.} However, for fainter targets, it may be necessary to stack multiple secondary eclipses to get a reasonable emission spectrum. For some of these cases, a band-integrated phase curve could still be obtained in a single visit. We suggest that in such cases, obtaining additional secondary eclipses (but not full phase curves, which would be much more time-intensive) is the best observing strategy. This will be particularly important for planets for which transmission spectroscopy has not been able to determine the composition of the atmosphere, for example due to high altitude aerosols (like GJ~1214b) or due to contamination from starspots, which generally has a lower impact on emission spectra.

\section{Conclusion}
\label{sec:conclusion}
We examined how photochemical hazes impact the global temperature structure, atmospheric circulation, phase curves and emission spectra of sub-Neptunes with a variety of atmospheric compositions and haze properties using the example of GJ~1214b. To do this, we utilized a set of simulations with a General Circulation Model (GCM) with wavelength-dependent radiative transfer. We assumed horizontally uniform hazes with a vertical profile derived from a microphysics model. Scattering and absorption from hazes are taken into account. We explored three different assumptions for the haze optical properties: soot, tholin, and ``maximally reflective'' hazes, which use the extinction cross section of soot but have a single-scattering albedo artificially set to 0.9999. We assumed that the haze vertical profile does not vary with atmospheric composition. To simulate adjusting the haze production rate, we scaled the optical depth by a constant factor.
Our main conclusions can be summarized as
\begin{itemize}
    \item The effect of photochemical hazes on the 3D temperature structure of sub-Neptunes strongly depends on the assumed haze optical properties. On the dayside, for the highly-absorbing soot hazes, the atmosphere heats up at low pressures compared to the clear-atmosphere case, while deeper layers experience strong cooling. The result is a strong thermal inversion. For the other extreme case we considered,``maximally reflective hazes", the deeper layers of atmosphere cool, while the temperature at low pressures remains unchanged. A smaller thermal inversion forms due to this cooling. For tholin hazes, which reflect a large fraction of the starlight but still absorb some of it, the changes are qualitatively between both of the other cases, with some warming at low pressures.
    \item The nightside temperatures are strongly determined by the amount of dayside cooling in deeper layers of the atmosphere. Thus, soot simulations tend to have the coldest nightsides at a given haze production rate.
    \item Photochemical hazes change the atmospheric circulation of sub-Neptunes. For high enough haze opacities, photochemical hazes tend to produce a single narrow equatorial jet, even for atmospheric compositions for which clear-atmosphere simulations predict a broad equatorial jet or multiple jets. The higher the haze production rate, the faster is the jet speed.
    \item Hazes tend to reduce the amplitude of absorption features on the dayside for moderate levels of haze production opacity and produce emission features for high haze production rates. With increasing metallicity, a higher haze opacity is needed for a thermal inversion to form and thus for emission features to appear in the dayside spectrum.
    \item Generally, phase curve amplitudes stay the same for low haze opacities and increase for higher haze opacities. Soot hazes result in larger increases to the phase curve amplitude than tholins or highly reflective hazes.
    \item Hazes significantly increase the scatter in the phase curve amplitude--mean molecular weight relationship. While the relationship generally holds true for hazy simulations if haze opacity and optical properties are held constant, neither haze production rates nor optical properties are currently known, and they may vary strongly between planets. It is thus no longer possible to estimate the mean molecular weight from the phase curve amplitude alone.
    \item However, when combining white-light phase curves with secondary eclipse spectra, the degeneracy between haziness and metallicity can be disentangled: If the haze abundance is high enough to affect the phase curve amplitude, the secondary eclipse spectrum will show signatures of hazes such as a close-to-isothermal temperature profile or emission features. Thus, phase curves continue to be a promising technique to characterize the atmospheres of sub-Neptunes for which transmission spectroscopy has proven to be challenging.
\end{itemize}
However, a few caveats remain: 
While our models qualitatively explain many features of the GJ~1214b phase curve (dayside flux, shape of nightside spectra, high phase curve amplitude), some tensions remain: Almost all models that match the white-light dayside flux show strong emission features (with one model showing subtle emission features), when the data shows subtle absorption features or blackbody spectra. Furthermore, none of our models can produce a nightside flux as low as in the Kempton reduction or a phase curve offset as large as in the Malsky reduction. It is likely that a more realistic haze model is needed to fully explain the observations (see discussion at the end of Section \ref{sec:gj1214b_comparison}).

To fully utilize the potential of phase curves in the study of sub-Neptunes, we will need robust measurements of the planet-to-star flux at better than 100 ppm precision.  This is particularly true for MIRI LRS, which covers the wavelength range where the emission from many hazy sub-Neptunes such as GJ 1214b peaks, and thus is the instrument of choice. However, MIRI data has the most complex instrument systematics of any JWST instrument. While data reduction for a relatively high signal-to-noise hot Jupiter phase curve led to consistent results across multiple teams \citep{BellEtAl2024WASP-43b}, for the more challenging case of GJ 1214b there are currently two independent analyses of the phase curve that do not agree, particularly on the nightside flux. We are still in the early days of JWST data analysis, and history has shown that for HST and Spitzer observations it took several years to build consensus on the best way to reduce and analyze the data. The high scientific potential of JWST phase curve observations motivates a continued push to improve our understanding of the instruments and their behavior in coming years.

\section*{Acknowledgments}
We thank Channon Visscher for providing equilibrium chemistry abundance tables. M.S. and M.Z. acknowledge support from the 51 Pegasi b Fellowship, funded by the Heising-Simons Foundation. This work is based in part on observations made with the NASA/ESA/CSA James Webb Space Telescope. These observations are associated with program \#1803. Support for this program was provided by NASA through a grant from the Space Telescope Science Institute, which is operated by the Association of Universities for Research in Astronomy, Inc., under NASA contract NAS 5-03127. RL was supported by NASA XRP grant 80NSSC22K0953 and STScI grant JWST-AR-01977.007-A. A portion of this research was carried out at the Jet Propulsion Laboratory, California Institute of Technology, under a contract with the National Aeronautics and Space Administration (80NM0018D0004).

%

\vspace{5mm}
\facilities{MPCDF, JWST}


\software{MITgcm \citep{AdcroftEtAl2004}, Numpy \citep{NumpyCitation}, SciPy \citep{SciPyCitation}, astropy \citep{Astropy2022}, matplotlib \citep{MatplotlibCitation}
          }



\appendix

\section{Haze optical properties}
\begin{figure*}
\begin{center}
\plotone{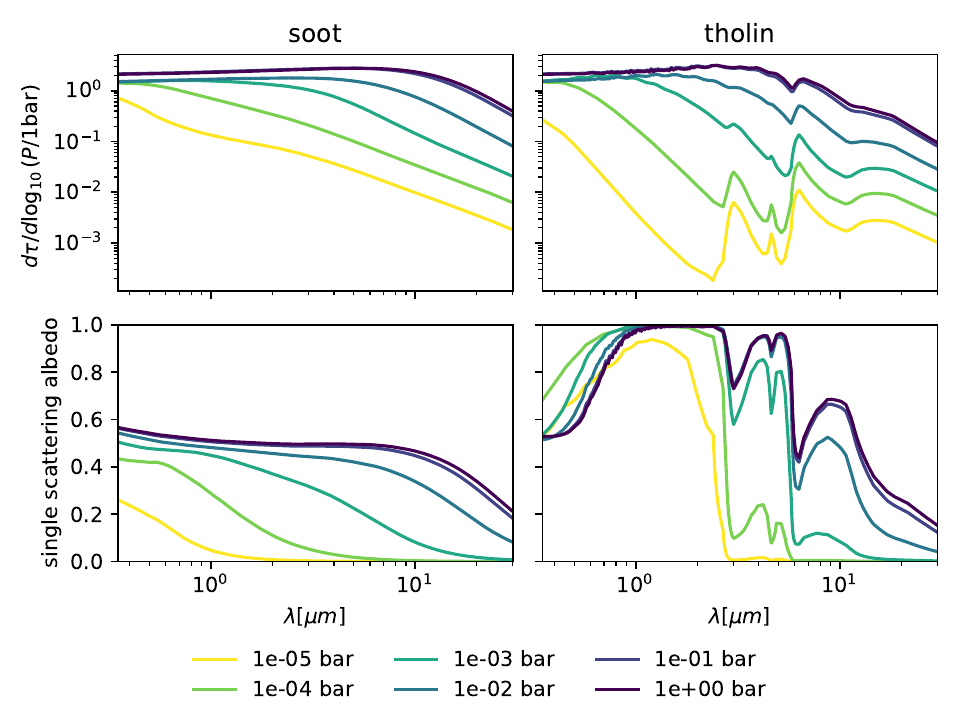}
\caption{Differential optical depth (as a proxy for extinction cross section, top row) and single scattering albedo (bottom row) for the assumed haze profile for soot (left) and tholin (right) hazes at different pressures.}
\label{fig:hazeprops_vs_wavelength}
\end{center}
\end{figure*}

The differential optical depth and single scattering albedos of the haze profiles we use are shown in Fig. \ref{fig:hazeprops_vs_wavelength}. An alternative visualization of the haze properties and in addition the asymmetry parameter can be found in Fig. 1 of \MalskyEtAlT.

\section{Contribution functions of clear-atmosphere simulations}
\begin{figure*}
\begin{center}
\plotone{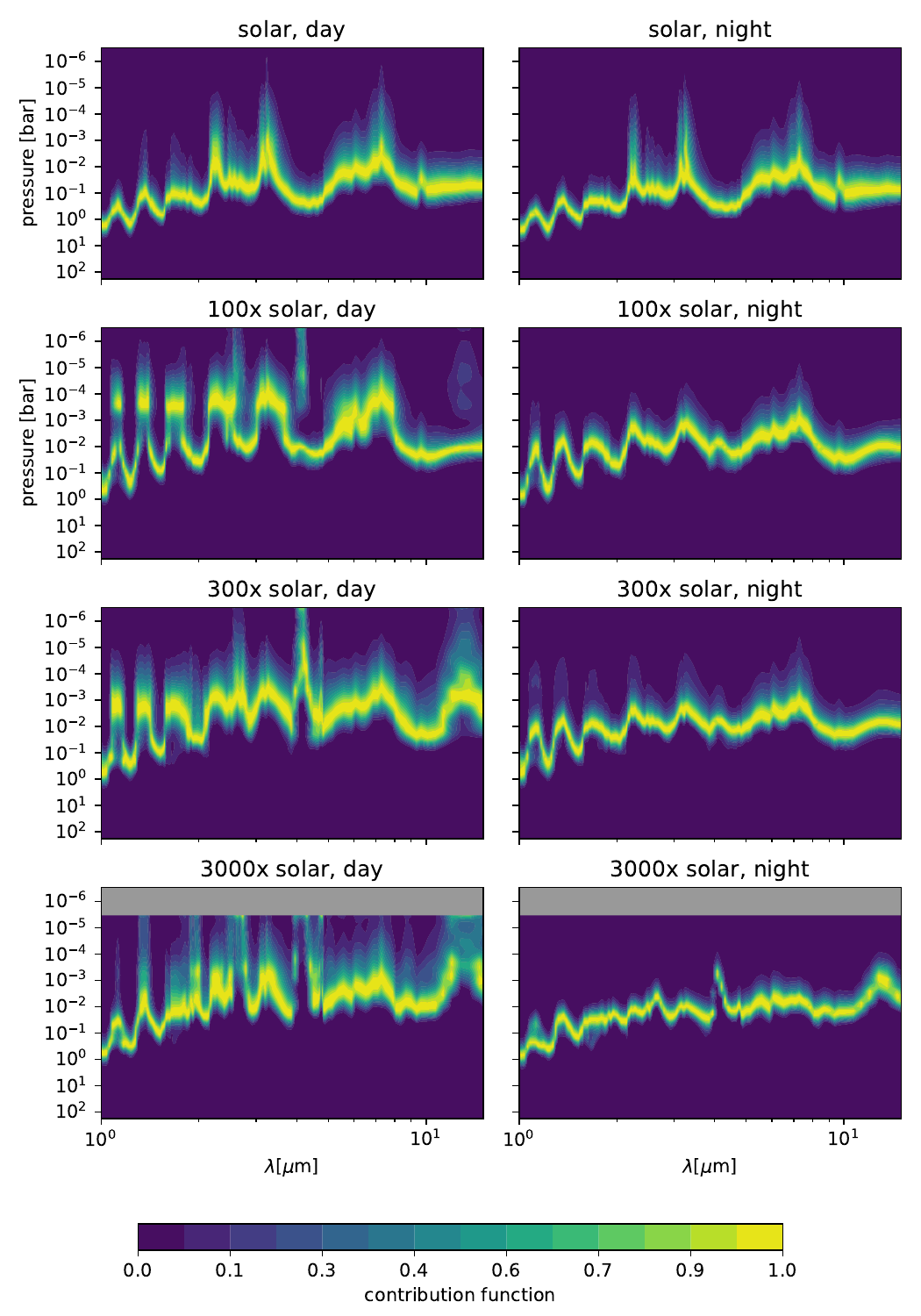}
\caption{Wavelength-dependent contribution functions for the dayside (left column) and nightside (right column) emission spectra. Shown are the clear-atmosphere simulations.}
\label{fig:contributionfunctions}
\end{center}
\end{figure*}

The contribution functions of the clear-atmosphere simulations are shown in Fig. \ref{fig:contributionfunctions}.

\section{Bins for spectral phase curve}
Due to the fairly low resolution of our post-processing (196 custom wavelength bins, with the correlated-k method being used to calculate the outgoing flux within each bin), the wavelength bins of the models do not line up exactly with the edges of the wavelength bins used in the observational spectral phase curves. For transparency, we show the wavelength bins used to generate Figures \ref{fig:spectralphasecurve_tholin} and \ref{fig:spectralphasecurve_maxrefl} in Table \ref{tab:wavelengthbins}. We found that the figures only changed marginally when choosing to display a slightly different set of the post-processing wavelength bins for display in the plots.



\begin{deluxetable}{rrrrr}




\tablecaption{Wavelength bins for spectral phase curves \label{tab:wavelengthbins}}


\tablehead{\colhead{plot title} & \colhead{$\lambda_{\textrm{obs},0}$} & \colhead{$\lambda_{\textrm{obs},1}$} & \colhead{$\lambda_{\textrm{model},0}$} & \colhead{$\lambda_{\textrm{model},1}$} \\ 
\colhead{($\mu$m)} & \colhead{($\mu$m)} &\colhead{($\mu$m)} & \colhead{($\mu$m)} & \colhead{($\mu$m)} } 

\startdata
5.25 & 5.0 & 5.5 & 5.08 & 5.45 \\
5.75 & 5.5 & 6.0 & 5.45 & 5.98 \\
6.25 & 6.0 & 6.5 & 5.98 & 6.45 \\
6.75 & 6.5 & 7.0 & 6.45 & 7.00 \\
7.25 & 7.0 & 7.5 & 7.00 & 7.62 \\
7.75 & 7.5 & 8.0 & 7.62 & 8.00 \\
8.25 & 8.0 & 8.5 & 8.00 & 8.54 \\
8.75 & 8.5 & 9.0 & 8.54 & 9.20 \\
9.25 & 9.0 & 9.5 & 9.20 & 9.50 \\
9.75 & 9.5 & 10.0 & 9.50 & 9.80 \\
10.25 & 10.0 & 10.5 & 9.80 & 10.40 \\
\enddata 




\end{deluxetable}

\bibliography{sample631}{}
\bibliographystyle{aasjournal}



\end{document}